\documentclass[aps,prd,10pt,twocolumn,
 superscriptaddress,floatfix,nofootinbib,
 amsmath,amssymb,altaffilletter,preprintnumbers]{revtex4-1}
\usepackage[colorlinks=true,pdfstartview=FitV,breaklinks=true]{hyperref}

\usepackage{physics}
\usepackage{bm}
\usepackage{graphicx}
\usepackage{multirow}
\usepackage{soul}
\usepackage{amsmath}
\usepackage{fontawesome}
\usepackage{mathrsfs}
\usepackage{amssymb}
\usepackage{yfonts}
\usepackage{makecell}
\usepackage{color}
\usepackage{xspace}
\usepackage{url}
\usepackage{verbatim}
\usepackage{mathtools}
\usepackage{upgreek}
\usepackage{units}
\usepackage{amstext}
\usepackage{times}
\usepackage{booktabs}
\usepackage{tabulary}
\usepackage{tabularx}
\usepackage{etoolbox}
\usepackage[version=4]{mhchem}
\usepackage[utf8]{inputenc}
\usepackage[dvipsnames,table]{xcolor}
\newcolumntype{Y}{>{\centering\arraybackslash}X}
\hypersetup{urlcolor=BlueViolet,
	    citecolor=Plum,
	    linkcolor=PineGreen}
\usepackage[official]{eurosym}
\definecolor{lightgray}{rgb}{0.9,0.9,0.9}	    
\definecolor{green}{rgb}{0,0.5,0}
\definecolor{red}{rgb}{1,0,0}
\definecolor{blue}{rgb}{0,0,0.5}

\long\def\symbolfootnote[#1]#2{\begingroup%
\def\thefootnote{\fnsymbol{footnote}}\footnotetext[#1]{#2}\footnotemark[#1]\endgroup}

\newcommand{\dbd}[2]{\ifmmode \frac{\textrm{d}#1}{\textrm{d}#2}\else $\textrm{d}#1/\textrm{d}#2$\fi}
\newcommand{\pbp}[2]{\ifmmode \frac{\partial#1}{\partial#2}\else $\partial#1/\partial#2$\fi}

\DeclareMathAlphabet{\mathpzc}{OT1}{pzc}{m}{it}
 \newcommand{\eV}{\text{e\kern-0.15ex V}\xspace}
 
 \newcommand{\MeV}{\text{M\eV}\xspace}
 \newcommand{\GeV}{\text{G\eV}\xspace}
 \newcommand{\TeV}{\text{T\kern-0.1ex \eV}\xspace}
 \newcommand{\cevns}{CE$\nu$NS\xspace}

\usepackage{tikz,xcolor,hyperref}

\definecolor{lime}{HTML}{A6CE39}
\DeclareRobustCommand{\orcidicon}{\hspace{-1mm}
	\begin{tikzpicture}
	\draw[lime, fill=lime] (0,0) 
	circle [radius=0.16] 
	node[white] {{\fontfamily{qag}\selectfont \tiny \,ID}};
	\draw[white, fill=white] (-0.0525,0.095) 
	circle [radius=0.007];
	\end{tikzpicture}
	\hspace{-3mm}
}

\foreach \x in {A, ..., Z}{\expandafter\xdef\csname orcid\x\endcsname{\noexpand\href{https://orcid.org/\csname orcidauthor\x\endcsname}
			{\noexpand\orcidicon}}
}

\begin{document}
\preprint{HRI-RECAPP-2023-13}

\title{The neutrino fog for dark matter-electron scattering experiments}

\author{Ben Carew}
\affiliation{School of Physics, The University of Sydney, and ARC Centre of Excellence for Dark Matter Particle Physics, NSW 2006, Camperdown, Sydney, Australia}

\author{Ashlee R.\ Caddell}
\affiliation{School of Mathematics and Physics, The University of Queensland, QLD 4072, Australia}

\author{Tarak Nath Maity}
\email{tarak.maity.physics@gmail.com}
\affiliation{Harish-Chandra Research Institute, A CI of Homi Bhabha National Institute, Chhatnag Road, Jhunsi, Prayagraj (Allahabad) 211019, India}
\affiliation{Regional Centre for Accelerator-based Particle Physics, Harish-Chandra Research Institute, Prayagraj (Allahabad) 211019, India}
\author{Ciaran A. J. O'Hare}\email{ciaran.ohare@sydney.edu.au}
\affiliation{School of Physics, The University of Sydney, and ARC Centre of Excellence for Dark Matter Particle Physics, NSW 2006, Camperdown, Sydney, Australia}


\smallskip
\begin{abstract}
The search for sub-GeV dark matter via scattering on electrons has ramped up in the last few years. Like in the case of dark matter scattering on nuclei, electron-recoil-based searches also face an ultimate background in the form of neutrinos. The so-called ``neutrino fog'' refers to the range of open dark-matter parameter space where the background of neutrinos can potentially prevent a conclusive discovery claim of a dark matter signal from being made. In this study, we map the neutrino fog for a range of electron recoil experiments based on silicon, germanium, xenon and argon targets. In analogy to the nuclear recoil case, we also calculate the ``edge'' to the neutrino fog, which can be used as a visual guide to where neutrinos become an important background---this boundary excludes some parts of the key theory milestones used to motivate these experiments.
\end{abstract}

\maketitle

\section{Introduction}\label{sec:intro}

Direct searches for dark matter (DM) in the low-mass (MeV--GeV) range have seen a surge in popularity over the last decade~\cite{Essig:2012yx,Essig:2011nj,Essig:2015cda,Graham:2012su,Battaglieri:2017aum,Essig:2022dfa,Kahn:2021ttr}. New technologies like the sensitive semiconductor-based charge-coupled devices (CCD)~\cite{Crisler:2018gci, SENSEI:2019ibb, SENSEI:2020dpa,DAMIC:2019dcn, DAMIC-M:2023gxo, DAMIC-M:2023hgj}, as well as previously established techniques like noble liquid time projection chambers~\cite{ZEPLIN-III:2011qer,XENON10:2011prx,XENON:2016jmt,XENON:2021qze,DarkSide:2018ppu,DarkSide:2022knj,XENON:2019gfn,PandaX-II:2021nsg} and cryogenic crystalline phonon detectors~\cite{EDELWEISS:2020fxc,SuperCDMS:2018mne,SuperCDMS:2020ymb} have all advanced to the stage where they can detect energy depositions at low enough energies to facilitate the identification of events ionising only a few primary electrons (and even down to a single electron~\cite{XENON:2021qze}). Energy thresholds at the level of tens of eV or less are required to extend the search into the ``light dark matter'' regime which, in the last decade or so, has simultaneously become increasingly interesting from a theory perspective. Many studies have revealed that there exists a range of simple and plausible cosmological scenarios that explain the observed DM abundance in the form of very light feebly-interacting particles---see e.g.~Refs.~\cite{Boehm:2003hm,Hall:2009bx,Chu:2011be,Boehm:2020wbt,Izaguirre:2015yja,Kuflik:2015isi,Dey:2016qgf, Kuflik:2017iqs, Dey:2018yjt, Maity:2019vbo, Dvorkin:2020xga,Elor:2021swj,Antel:2023hkf,DelaTorreLuque:2023olp}---whilst escaping naive bounds on the lightest mass allowed for a thermal relic DM particle~\cite{Hut:1977zn,Lee:1977ua}. 

\begin{figure*}
\begin{center}
\includegraphics[trim = 0mm 0mm 0mm 0mm, clip, width=0.49\textwidth]{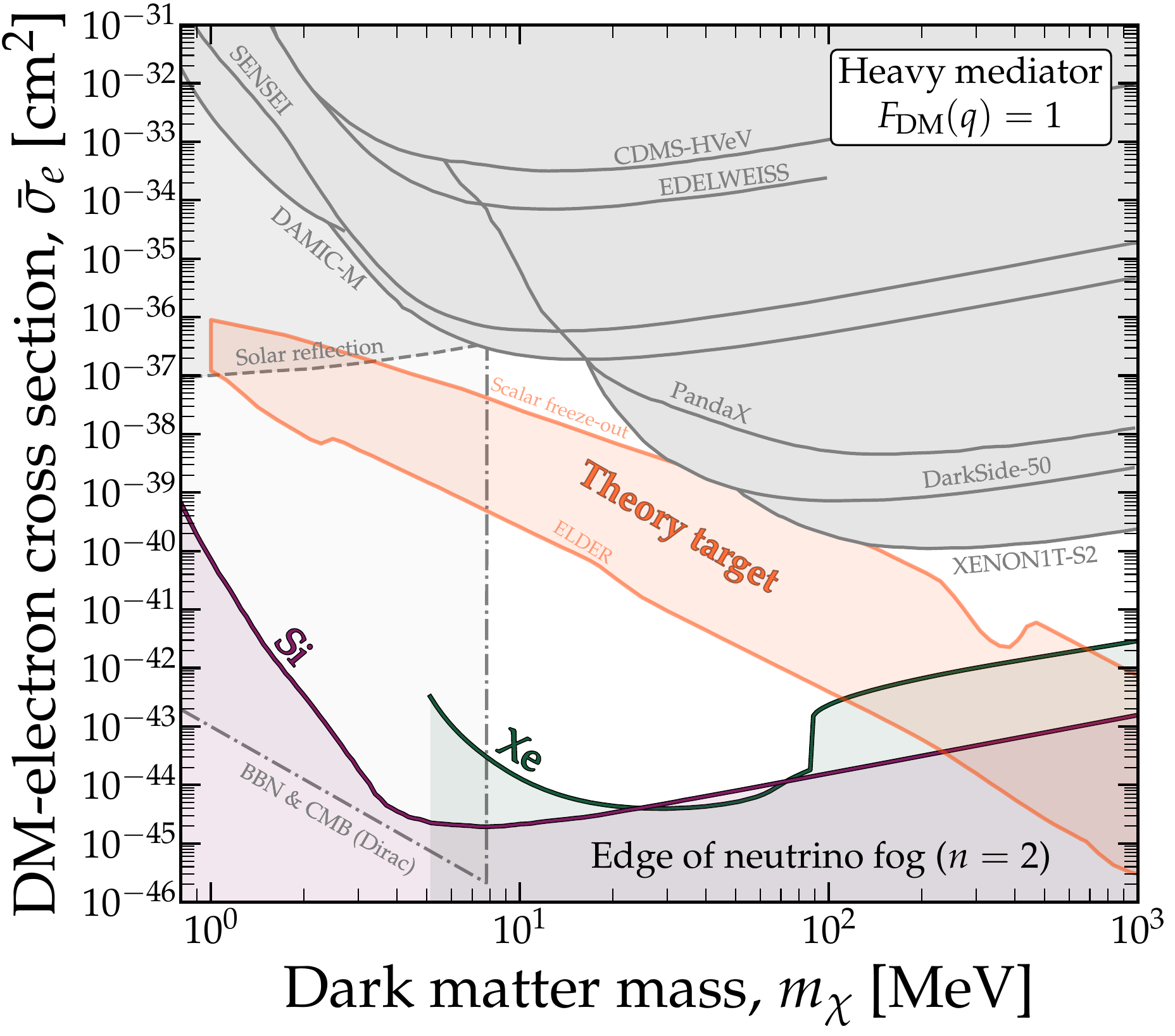}
\includegraphics[trim = 0mm 0mm 0mm 0mm, clip, width=0.49\textwidth]{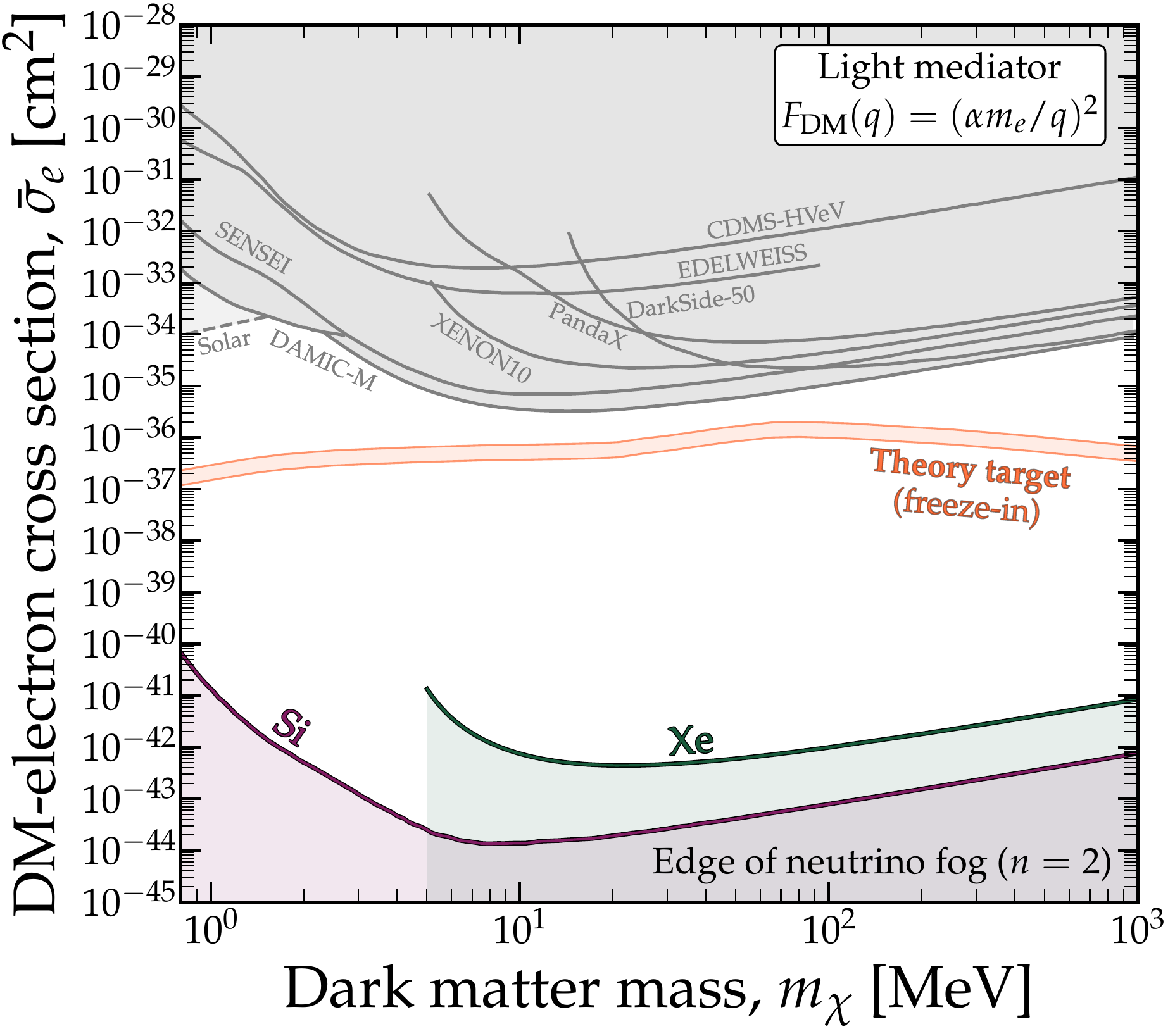}
\caption{Summary of the main result obtained in this paper---the neutrino fog ``edges'' for silicon and liquid xenon (LXe) experiments for the commonly-used heavy (left) and light (right) DM-electron interaction models. We define the ``edge of the neutrino fog'' as the cross section at which the scaling of a DM discovery limit scales as $\sigma \propto (MT)^{-1/n}$ with $n>2$---implying slower scaling than the Poisson expectation (full visualisations of the value of $n$ as a function of both $m_\chi$ and $\bar{\sigma}_e$ are shown in Fig.~\ref{fig:nufogs}). To provide additional context, we have shown the community's key theory milestones for each case, i.e. models that explain the DM abundance whilst having the possibility of DM-electron scattering. These correspond to the window between a simple scalar-DM freeze-out scenario~\cite{Essig:2017kqs, Boehm:2003ha, Boehm:2003hm, Izaguirre:2015yja, Essig:2022dfa} and the ELDER model~\cite{Kuflik:2015isi, Kuflik:2017iqs} for the heavy mediator case; and for the light mediator case, the DM freeze-in scenario~\cite{Essig:2012yx,Essig:2022dfa}. Constraints from existing experiments are shown in gray: SENSEI~\cite{SENSEI:2020dpa}, DAMIC-M~\cite{DAMIC-M:2023gxo, DAMIC-M:2023hgj}, CDMS-HVeV~\cite{SuperCDMS:2018mne,SuperCDMS:2020ymb}, EDELWEISS~\cite{EDELWEISS:2020fxc}, PandaX~\cite{PandaX-II:2021nsg}, DarkSide-50~\cite{DarkSide:2022knj}, XENON10~\cite{Essig:2012yx}, XENON1T-S2~\cite{XENON:2019gfn}, and XENON1T for solar-reflected DM~\cite{An:2017ojc,An:2021qdl,Emken:2021lgc}. Since big-bang nucleosynthesis (BBN) and the cosmic microwave background (CMB) observations are often incompatible with very low-mass DM, we have shown an indicative example of such a bound under the assumption that DM is a Dirac fermion with a mediator three times its mass~\cite{Giovanetti:2021izc}.} 
\label{fig:Combined}
\end{center}
\end{figure*}

In order to fully test these models, novel detectors exploiting DM-electron recoils will continue to grow in size, reaching the 10--100~kg scale in the case of futuristic proposals for solid-state experiments~\cite{Oscura:2022vmi,TESSERACT} and potentially up to the 100 ton-year-scale for the ultimate liquid-xenon ``observatory''~\cite{Akerib:2022ort}. This will inevitably come with challenges associated with tackling problematic backgrounds and limited event discrimination, both of which will inhibit their discovery potential for DM models that generate events at very low energies. Assuming that these experiments will continue, however, one thing is certain, and that is that the neutrino background is on the horizon. It was foreseen long ago that low-background nuclear-recoil detectors could eventually reach sensitivity sufficient to detect neutrinos from astrophysical sources if DM was not detected sooner~\cite{Monroe:2007xp,Vergados:2008jp,Strigari:2009bq,Gutlein:2010tq}. DM remains undetected, and still experiments proceed, so it seems as though this situation is almost upon us. Indeed, multi-ton-scale liquid xenon experiments are now poised to make the first detection of solar neutrinos via coherent elastic neutrino-nucleus scattering (\cevns) towards the end of their planned data-taking campaigns---see e.g.~Refs.~\cite{XENON:2020gfr,Xiang:2023csc} for some further discussion and potential obstacles to this anticipated discovery. 

Neutrinos are the ultimate background for DM experiments utilising signatures arising from recoils~\cite{Billard:2013qya}. But as well as being the final background, they are also foreseen to be a problematic one. Due to a numerical coincidence between the typical nuclear recoil energies generated by the very light but relativistic solar neutrinos and the heavy but non-relativistic dark matter, the event rate spectra of these two signals in the nuclear recoil channel happens to line up very closely. In fact, for some DM masses and some sources of solar neutrino, the two signals are almost identical~\cite{Billard:2013qya}. This leads to the now widely-known concept of the ``neutrino floor''---the region of parameter space where the presence of the neutrino background could prevent the conclusive identification of a DM signal, see Refs.~\cite{OHare:2016pjy,Dent:2016iht, Dent:2016wor,AristizabalSierra:2017joc,Gonzalez-Garcia:2018dep,Papoulias:2018uzy,Boehm:2018sux,Nikolic:2020fom,Munoz:2021sad,Calabrese:2021zfq,Sierra:2021axk,Tang:2023xub,Herrera:2023xun} for many previous studies on different aspects of this concept. More precisely, the central problem is that the neutrino background must be modelled with the finite systematic uncertainty associated with our incomplete knowledge of it, and this turns out to be significant given the low numbers of events we expect from neutrinos. A sensitivity ``floor'' then arises because putative signal events coming from DM are indistinguishable from the expected fluctuations in the background rate. 

However, as pointed out in previous studies~\cite{Ruppin:2014bra,Gelmini:2018ogy,Sassi:2021umf,Gaspert:2021gyj,OHare:2020lva}, and then formalised in Ref.~\cite{OHare:2021utq}, the neutrino background and the DM signals are never exactly the same, and so there can never be a hard sensitivity ``floor''. This has inspired the recent re-branding of the situation in terms of a neutrino ``fog''. Once an experiment reaches the neutrino fog, further progress through the parameter space is slowed but not halted entirely. Extending the metaphor further, the ``opacity'' of the neutrino fog can be quantified in terms of how slowly an experiment progresses through the parameter space---doing so reveals the specific regions of parameter space that are more difficult to push through due to the presence of specific overlapping neutrino backgrounds that confound the positive discovery of DM. The neutrino fog has been mapped for nuclear-recoil experiments in the $\gtrsim$ GeV mass range~\cite{OHare:2021utq}, so our goal here is to do the same for electron recoil experiments in the sub-GeV mass regime.

Compared to nuclear recoils, sub-GeV DM-electron recoils exist at extremely low energies and, therefore, require specific experimental setups to identify. A range of experiments are being developed right now, which form the inspiration for the simplified detector cases we consider here. We focus on two different classes of low-background electron-recoil searches that make use of solid-state semiconductor targets and liquid noble gases as these set the leading limits\footnote{There are also interesting upcoming experiments based on gas targets\,\cite{Gerbier:2014jwa, Hamaide:2021hlp, NEWS-G:2023qwh} which we defer to future studies.}. For the former, we consider both silicon and germanium-based detectors, as is the case for SENSEI~\cite{Crisler:2018gci,SENSEI:2019ibb,SENSEI:2020dpa}, DAMIC~\cite{DAMIC:2019dcn,DAMIC-M:2023gxo}, and their potential successor Oscura~\cite{Oscura:2022vmi} utilising silicon skipper-CCDs, as well as crystalline phonon detectors like SuperCDMS~\cite{SuperCDMS:2018mne,SuperCDMS:2020ymb}, EDELWEISS~\cite{EDELWEISS:2020fxc} and TESSERACT~\cite{TESSERACT}. For noble-liquid targets, we consider the examples of xenon dual-phase time projection chambers (TPCs) such as LZ~\cite{LZ:2021xov,LZ:2022ufs}, XENONnT~\cite{XENON:2023sxq} and PandaX~\cite{PandaX-4T:2021bab}, and their potential successors DARWIN~\cite{Aalbers:2016jon} and XLZD~\cite{Aalbers:2022dzr}; as well as argon-based detectors like DarkSide~\cite{DarkSide-50:2023fcw}. Work on the design and construction of next-generation detectors is already underway, and the community's plan appears to be to, at the very least, reach some proposed theory ``milestones''. These consist of parts of the parameter space of DM-electron cross sections and DM masses that can be predicted by minimal models which correctly satisfy the observed cosmological abundance of DM. To see some examples in context with the current and near-future experimental landscape, we refer to the recent Snowmass white paper on the subject~\cite{Essig:2022dfa}, as well as those included in Fig.~\ref{fig:Combined} which shows the main result of this paper.

Our work builds on two previous studies of the DM-electron neutrino floor, Ref.\,\cite{Essig:2018tss} and Ref.\,\cite{Wyenberg:2018eyv}. The former evaluated the effects of the neutrino background on the projected sensitivity of some specific detector models, based on Si and Ge semiconductor devices, and liquid xenon. That study revealed the importance of considering the misidentified neutrino-nucleus recoils which can also generate few-electron ionisation signatures that are challenging to distinguish from true electron recoil events---we also incorporate this lack of knowledge into our analysis using a similar model for the ionisation yield from nuclear recoils. The second of those two studies involved similar calculations for sub-\GeV DM detection using a germanium target, while including the effects of uncertainties in the DM velocity distribution, and the ability of some detectors to discriminate between electron and nuclear recoils on the discovery limits. We build on these previous works by using recent state-of-the-art methods of computing the DM-electron scattering event rate, and instead mapping the whole of the continuous neutrino \textit{fog} using the methodology of Ref.~\cite{OHare:2021utq}.

The paper is structured as follows: In Sec.~\ref{sec:exp} we introduce the different types of experiments whose corresponding neutrino fogs we will calculate.. Then, in Sec.~\ref{sec:rates}, we outline the main ingredients for our DM and neutrino event rate calculations observable by those experiments. Finally, in Sec.~\ref{sec:results}, we define the concept of the neutrino fog and present our results, before concluding in Sec.\,\ref{sec:conc}.

\section{Experiments}\label{sec:exp}
The solid-state detector, DAMIC-M \cite{DAMIC-M:2023gxo, DAMIC-M:2023hgj} currently sets the strongest limits on DM-electron cross sections for DM masses below $15.1\,$\MeV\footnote{This is the case for models where DM interacts via a heavy mediator, and is true for the full mass range in the light mediator case. We introduce the distinction between them in Sec.~\ref{sec:rates}.}, superseding SENSEI at the time of writing. SENSEI currently uses a single 2~g silicon skipper-CCD, with plans to expand to 100~g~\cite{SENSEI:2020dpa}. The DAMIC-M limit~\cite{DAMIC-M:2023gxo} was set with 85.23 g-days of data, whereas their recent daily modulation-based search \cite{DAMIC-M:2023hgj} used 39.97 g-days. The joint proposal of these two collaborations is Oscura \cite{Oscura:2022vmi}, which is planned to be a 10~kg silicon skipper-CCD experiment that will test most of the well-motivated regions of the parameter space, displayed in Fig.~\ref{fig:Combined}. Germanium is also a common target choice, used in experiments such as SuperCDMS and EDELWEISS. The small semi-conductive band gap in these targets means detectors can probe recoil energies down to $\approx$ 3.6 \eV for silicon and 2.9 \eV for germanium. Further skipper-CCDs have improved sensitivity to single-electrons by allowing repeated charge readouts from each pixel~\cite{Tiffenberg:2017aac}.

The leading electron-recoil limits for the heavy-mediator case $F_{\rm DM}=1$ in the $>$15.1 \MeV region come from liquid noble dual-phase TPCs, which have grown in target mass from a few kilograms to a few tonnes in the last two decades. DarkSide uses liquid argon (LAr), while XENONnT, PandaX-4T and LZ use liquid xenon (LXe). These detectors are capable of measuring two quantities per event: a prompt scintillation signal (S1) emitted at the interaction site, and a delayed ionisation signal (S2) when the drifted charge arrives at the gas phase. The lowest energy thresholds are reached using an S2-only analysis. TPCs have the ability to differentiate electron and nuclear recoils by measuring the ratio of the signals $S1/S2$, but by using only the S2 signal, this discrimination is sacrificed in favour of considering a larger number of events at very low energies that only have a measurable S2. Despite this, liquid noble TPCs have still achieved the lowest observed background rate per unit mass, allowing them to lead the search for DM across most of the mass range to date. The XLZD project aims at a tenfold increase in exposure over current detectors, with forty to sixty tons of active target mass, potentially reaching ultimate exposures on the order of 100 ton-years or more~\cite{Aalbers:2022dzr}.

The information about the recoiling electron that is accessible to each of these types of experiments will be different, and there are many experimental details that are not necessarily well defined at present to allow us to create an exhaustive model of each detector. Instead, we make some broad assumptions that will allow us to illustrate the important effects of the neutrino background on the discovery of DM signal---similar to the simplifying assumptions used in other studies of the neutrino floor. One aspect that will become important later is the fact that most analyses of DM-electron scattering work by measuring discrete numbers of ionised electrons, $n_e$, which are functions of the initial recoil energy of the electron, $E_{e}$. This is signal that we will assume is observable in this study. Examples of these signals are shown in the left-hand panels of Fig.~\ref{fig:nufogs}, and will be derived in the next section.

So in light of the community's current plans, we will map the neutrino fogs for Si, Ge, LAr, and LXe targets. As with all previous studies of the neutrino floor and neutrino fog in other contexts, we assume that the detectors are idealised in all ways other than the factors that are non-negotiable from basic physics. Most importantly, we impose a hard single-electron cut-off so as to not incorporate DM masses that are not kinematically able to ionise even one electron. On the other hand, we will assume that there are no important \textit{non}-neutrino backgrounds. Given that these so-called ``ultimate'' detectors are pitched as being limited solely by the neutrino background, this is a reasonable assumption here. Of course, our detector models will still be somewhat idealistic, but the motivation behind this choice is to map the best-case scenario neutrino fog, i.e.~the parameter space where the neutrino background is guaranteed to be important, regardless of other experimental specifics which may evolve as technology improves. 

The most important factor for our detector models turns out to be the \textit{nuclear}-recoil background due to \cevns. Given that in the energy range of interest, the neutrino-nucleus recoil events generated by the low-energy solar neutrino fluxes will dominate the neutrino background, these events do make a substantial impact on the shape of and features within the neutrino fog, so we must understand how nuclear recoil events manifest inside experiments that are nominally searching for electron recoils. To incorporate them, we need a model for the quenching factor or ionisation yield, i.e. how much ionisation energy is observed for a given initial nuclear recoil energy. This ionisation yield, $Y$, is a function of energy, where the electron equivalent energy, $E_e$, is given in terms of the nuclear recoil energy $E_{r}$ as $E_e = Y(E_{r})\times E_{r}$. The exact nature of this function for different detectors is not always determined in full---in particular, its shape towards low energies is a subject of active investigation by experimental collaborations, see e.g.~the recent SuperCDMS measurement at 100~\eV \cite{SuperCDMS:2023geu} in silicon. So we will not attempt an exhaustive study of different ionisation yield models here, which in any case was done already in Ref.~\cite{Essig:2018tss}. Instead, we take the baseline or fiducial models for xenon, silicon and germanium outlined in Ref.~\cite{Essig:2018tss}, which are reasonable estimates without being overly optimistic with regard to their very-low energy sensitivity. These are based on the well-known Lindhard model, which is fit to data towards higher energies, and then given a hard cutoff at low energies based on an extrapolation from that higher-energy data. 

\section{Rates}\label{sec:rates}

\subsection{Dark matter-electron scattering}
DM models with sub-\GeV masses are often too light to deposit sufficient energy to generate detectable nuclear recoil signals, so the more interesting alternative signal for DM detection in this mass range is the measurement of the electrons ionised from their atoms in direct DM-electron scattering events\,\cite{Essig:2011nj,Essig:2012yx}. An electron recoil occurs when a particle interacts with an atomic electron in the medium. If the particle deposits an energy $E_{\text{dep}} > {I_{njl}}$---where $I_{njl}$ is the ionisation energy for the shell with the principal quantum number $n$, total and orbital angular momentum quantum numbers $j$ and $l$--- then the electron will be ionised with energy $E_e = E_{\text{dep}}-{I_{njl}}$, leading to a detectable signal.

To obtain accurate event rates that are aligned with the current state-of-the-art in the field, we make use of two recent codes for each type of experiment: DarkELF~\cite{Knapen:2021bwg}\footnote{The utilization of other codes, such as EXCEED-DM\,\cite{Trickle:2022fwt} or QEdark\,\cite{Essig:2015cda}, does not appreciably alter our results, as shown in appendix\,\ref{app:other_codes}.} for solid-state targets, and AMPSCI~\cite{Caddell:2023zsw, ampsci}  for liquid xenon and argon. We refer the reader to each of those original references for a full discussion of the event rate calculations and their associated assumptions and uncertainties. That being said, we will also briefly sketch the main ingredients of a DM-electron scattering rate calculation so as to aid discussion later.

One must begin with an atomic or molecular form factor $K$ to calculate the DM electron recoil event rates. This is a function of electron recoil energy and momentum transfer that characterises the probability that an electron sitting in a particular atomic orbital will get ionised from the atom. The total atomic form factor is the sum of the contribution from each shell, weighted by the shell occupancy, expressed as a matrix element,
\begin{align}
    K_{njl}(E_e,q) &= \sum_m \sum_f \left|\bra{f}e^{i q\cdot r}\ket{njlm}\right|^2 E_H\rho_f(E_e),
\end{align}
where $q$ is the momentum transfer, $\rho_f(E_e)$ is the density of final (ionised) states, and $E_H=m_ec^2\alpha^2\approx27\,{\rm eV}$ is the Hartree energy, introduced to make $K$ dimensionless~\cite{Roberts:2019chv}. This quantity is non-trivial to calculate for two reasons. Firstly, for high values of $q$, the atomic electron wavefunctions $\ket{njlm}$ must be calculated relativistically~\cite{Roberts:2015lga}. Secondly, the ionised electron wavefunction $\bra{f}$ cannot be accurately approximated as a plane wave due to the continued influence of the atomic cloud from which it has been ionised~\cite{Roberts:2019chv}.
In practice, energy eigenstates should be used for the final state. These are typically normalised on the energy scale, with $\rho_f$ included in the definition of the wavefunction~\cite{Caddell:2023zsw}. In the case of liquid xenon and argon targets, we source only the ionisation form factors and use them to compute the event rates as detailed below. The package AMPSCI~\cite{Caddell:2023zsw, ampsci} computes the form factor using a relativistic Hartree-Fock approximation accounting for many-body effects.

With the atomic form factor in hand, this can then be integrated over the kinematically allowed range for the momentum transfer to the electron, as well as over DM velocity distribution, $f(v)$, to form the velocity-averaged cross-section:
\begin{align}
  \dv{\expval{\sigma v}}{E_e} = \frac{\bar{\sigma}_e}{2m_e}\int\mathrm{d}v\frac{f(v)}{v}\int_{q_-}^{q_+}a_0^2 q \mathrm{d}q\abs{F_{\rm DM}(q)}^2K_{njl}(E_e,q) \, .
\end{align}
The momentum limits depend on energy and the DM speed as,
\begin{align}
q_{\pm} &= m_\chi v \pm \sqrt{m_\chi^2 v^2 - 2m_\chi E_{\rm dep}} \, ,
\end{align}
where $a_0 = 1/m_e\alpha$ is the Bohr radius, $m_\chi$ is the DM mass, and $\bar{\sigma}_e$ is a reference scattering cross-section between DM and a free electron defined at a momentum transfer of $q = \alpha m_e$~\cite{Essig:2011nj}. The function $F_{\rm DM}(q)$ is the so-called ``DM form factor'' which is used to encapsulate the momentum dependence of the DM-electron cross section away from the reference value, which depends on the mass of the mediator particle for the interaction ---this is discussed further below. The integral over velocity includes $f(v)$, the DM velocity distribution. We adopt the truncated Gaussian distribution of velocities that arises under the Standard Halo Model assumption for the shape of the Milky Way's DM halo\footnote{We do not explore alternative halo models here. See e.g.~Refs.\,\cite{Buch:2020xyt, Radick:2020qip, Maity:2020wic, Maity:2022enp} for the effects of non-standard velocity distributions in the case of DM-electron scattering experiments.} with recommended parameter values for the width of the distribution, the escape speed and Earth's velocity~\cite{Evans:2018bqy}. The differential event rate per unit detector mass (assuming a target nucleus of mass $m_A$) is then simply written as,
\begin{align}
  \dv{R_{\chi}}{E_e} &= \frac{1}{m_A}\frac{\rho_{\chi}}{m_\chi}\dv{\expval{\sigma v}}{E_e} \, ,
\end{align}
where $\rho_\chi= 0.3$ \GeV/cm$^3$ is the value of the local DM density adopted by convention.

The so-called ``DM form factor'', $F(q)$, is a function of momentum transfer that depends on the mass of the mediator particle involved in the DM-electron interaction, $m_\phi$:
\begin{equation}
F_{\rm D M}(q)=\frac{\alpha^2 m_e^2+m_\phi^2}{q^2+m_\phi^2} \, .
\end{equation}
We adopt the usual convention in the literature, which is to work in one of two regimes: where the mediator mass is either heavy, in which case $F_{\rm DM}(q) = 1$, or light \mbox{$F_{\rm DM}(q) = (\alpha m_e /q)^2$}. 

The atomic form factor is the most demanding aspect of the event rate calculation and has the largest associated theoretical uncertainties. For semiconductor targets, the event rates calculated using DarkELF are based on the energy-loss function--an approach that leverages experimental data on the targets of interest. It also accounts for in-medium screening effects, which can have a sizeable impact on the measurable rate. We use the result from the ``GPAW'' calculation of this which is based on time-dependent density functional theory, outlined in Refs.~\cite{Knapen:2021run, Knapen:2021bwg}. 

Once the differential event rate has been calculated, we then translate this into the observable signal by converting $E_e$ into a discrete number of ionised electrons, $n_e$, with the use of an ionisation yield function. For LXe we use the formulae presented in Ref.\,\cite{Essig:2017kqs}, whereas for LAr we refer to Ref.\,\cite{DarkSide-50:2023fcw} with electron recoil ionisation yield given in Ref.\,\cite{DarkSide:2021bnz}. For semiconductor targets, the electron recoil energy is instead converted into the number of electrons excited above the material bandgap ($E_{\rm gap}$): 
\begin{equation}
n_e = 1 + {\rm floor} \left[ \frac{E_e - E_{\rm gap}}{\epsilon}\right] \, ,
\label{eq:EetoneSolid}
\end{equation}
where for silicon $E_{\rm gap} = 1.11$\,\eV and $\epsilon = 3.6$\,\eV and for germanium $E_{\rm gap} = 0.67$\,\eV and $\epsilon = 2.9$\,\eV. Some example DM-induced event rates for Si and LXe are shown in the left-hand panels of Fig.\,\ref{fig:nufogs} for a few benchmark DM models.

\subsection{Neutrino scattering rate}\label{sec:nu}

Any neutrinos reaching the detector could have been produced by various sources, including the sun, the earth, the atmosphere, and nuclear reactors. For the case of electron-recoil DM searches, only solar neutrinos are a relevant background\,\cite{Essig:2018tss, Wyenberg:2018eyv}. We use values for the solar neutrino flux normalisations and uncertainties from Ref.~\cite{OHare:2020lva}. The fluxes and theoretical systematic uncertainties are primarily calculated under the high-metallicity Standard Solar Model (SSM), with the exception of $^8$B, for which experimental data provides a smaller uncertainty~\cite{Bergstrom:2016cbh, Super-Kamiokande:2001ljr,Borexino:2008fkj,Borexino:2017uhp,Super-Kamiokande:2010tar,KamLAND:2011fld,SNO:2011hxd,SNO:2018fch,Gonzalez-Garcia:2023kva}.

As mentioned in Sec.~\ref{sec:exp}, while neutrinos will scatter on both the electrons and the nucleons in the target medium, for DM searches based on very low-energy electron recoil events, the misidentified electron recoils from \cevns turn out to create the dominant neutrino background~\cite{Essig:2018tss}. This is because the nuclear recoil rates are several orders of magnitude larger than electron recoils for neutrinos, in our energy range of interest. For the neutrino-nucleus interactions, we use the differential cross-section for \cevns~\cite{Freedman:1973yd,Freedman:1977,Drukier:1983gj} (only measured by COHERENT \cite{COHERENT:2017ipa, COHERENT:2020ybo,COHERENT:2021xmm}) with respect to the nuclear recoil energy $E_{r}$, given by
\begin{align}
  \dv{\sigma}{E_{r}} &= \frac{G_F^2}{4\pi}Q_W^2 m_N \left(1-\frac{m_N E_{r}}{2E_\nu^2}\right)F^2(E_{r}) \, ,
\end{align}
with neutrino energy $E_\nu$, the mass of the nucleus $m_N$, the Fermi coupling constant $G_F = 1.166\times 10^{-5}$\,\GeV$^{-2}$, and $F(E_{r})$ is the Helm form factor. The weak nuclear hypercharge of a nucleus containing $N$ neutrons and $Z$ protons is denoted by $Q_W = N - Z(1-4\sin^2\theta_W)$ with $\sin^2\theta_W=0.2387$\,\cite{Erler:2004in}. The effect of uncertainties related to the Weinberg angle and nuclear form factor is expected to be insignificant\,\cite{AristizabalSierra:2021kht}. The differential scattering rate is then,
\begin{align}
    \dv{R_{\nu}}{E_{r}} &= N_T \int_{E_{\nu}^{\text{min}}}\dv{\sigma}{E_{r}}\dv{N_\nu}{E_\nu}\mathrm{d}E_\nu \, ,
\end{align}
where $N_T$ is the number of target atoms in the detector and ${{\rm d} N_\nu}/{{\rm d}E_{\nu}}$ is the differential neutrino flux. The integral is performed over all neutrino energies high enough to produce a nuclear recoil of energy $E_{r}$,
\begin{align}
    E_\nu^{\text{min}} &= \sqrt{\frac{m_N E_{r}}{2}} \, .
\end{align}

As we did in the electron recoil case above, we must then convert these recoil energies into the detectable signal. For LXe we convert $E_{r}$ into a discrete number of ionised electrons again using the formalism of Ref.\,\cite{Essig:2018tss} adopting their ``fiducial'' model. For LAr we again follow Ref.\,\cite{DarkSide-50:2023fcw} with a nuclear recoil ionisation yield from Ref.~\cite{DarkSide:2021bnz}. In case of semiconductors, we convert nuclear recoil energy $E_{r}$ to ionised electron energy $E_e$ by making the following transformation\,\cite{Essig:2018tss},
\begin{align}
    \dv{R_{\nu}}{E_e} &= \dv{R_{\nu}}{E_{r}} \times \frac{1}{Y(E_{r}) + E_{r} \dv{Y(E_{r})}{E_{r}}} \, ,
\end{align}
where $Y(E_{r})$ is the ionisation yield discussed in the previous section. After this is applied, $E_e$ is then converted to $n_e$ using Eq.\eqref{eq:EetoneSolid} and so can be readily compared to the DM event rates we calculated before. We show the summed neutrino event rates as the dashed grey lines in the left-hand panels of Fig.\,\ref{fig:nufogs}.

\section{Results}\label{sec:results}

\begin{figure*}
\begin{center}
\includegraphics[trim = 0mm 0mm 0mm 0mm, clip, width=0.42\textwidth]{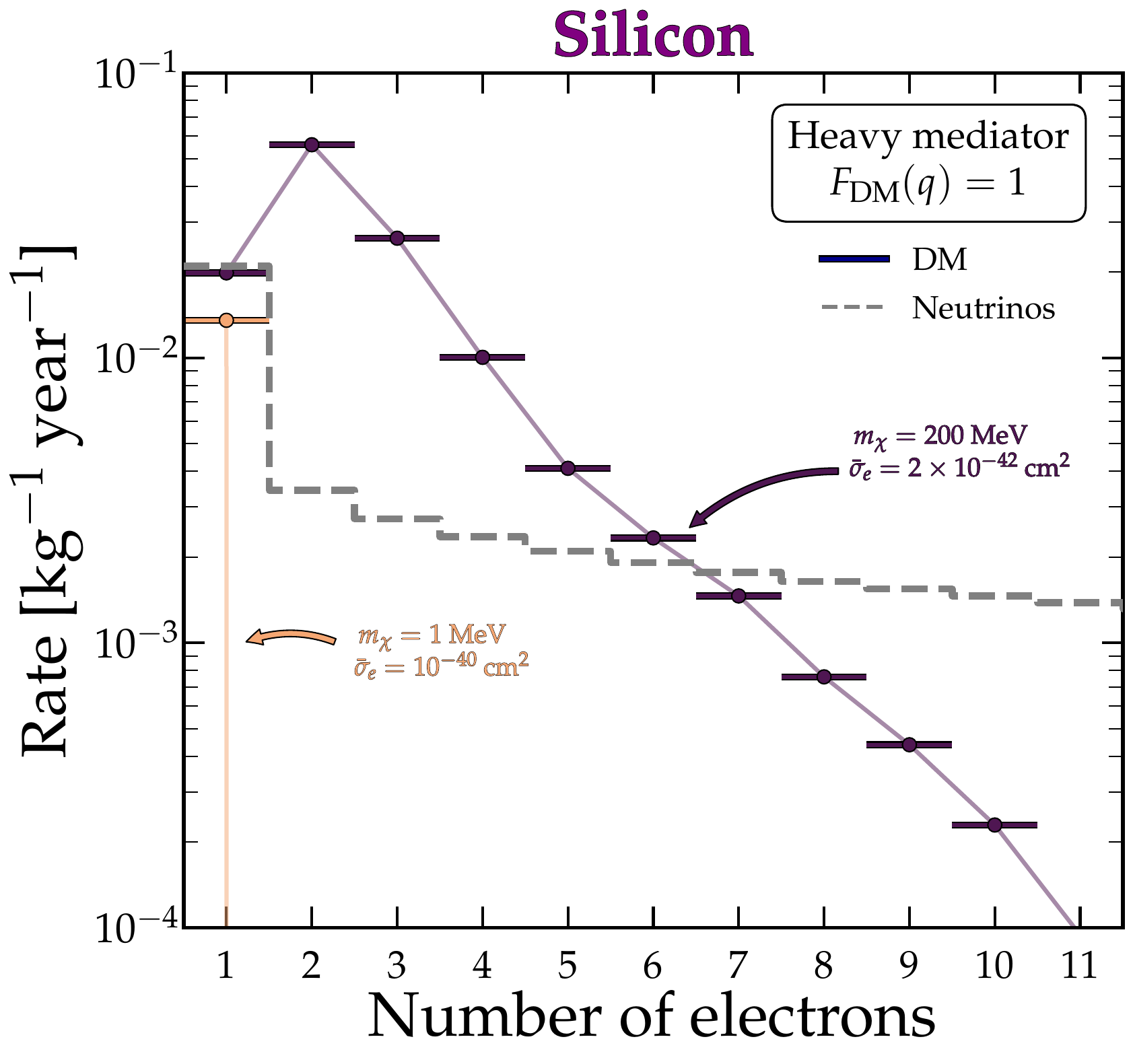}\phantom{0000}
\includegraphics[trim = 0mm 0mm 0mm 0mm, clip, width=0.49\textwidth]{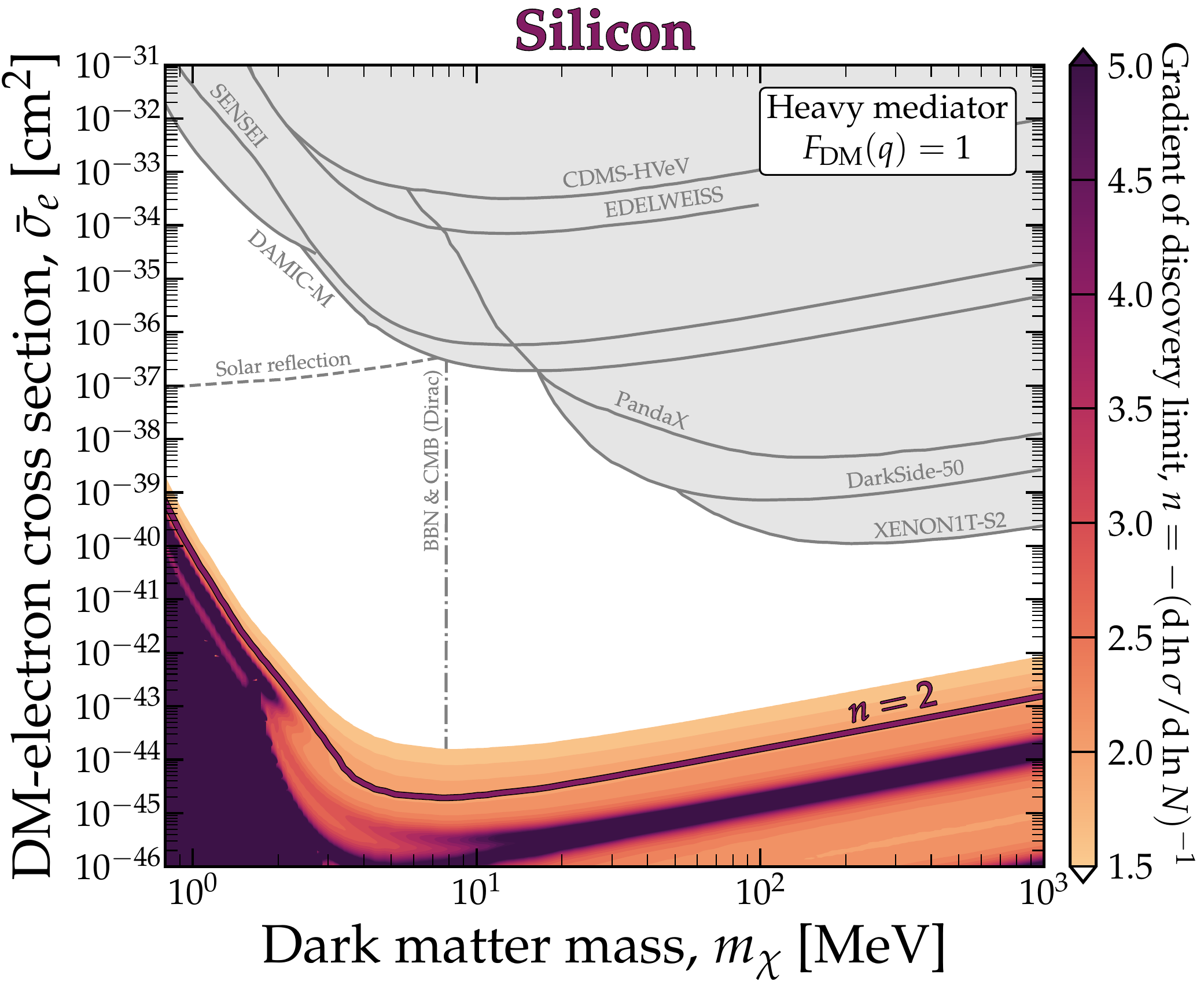}

\includegraphics[trim = 0mm 0mm 0mm 0mm, clip, width=0.42\textwidth]{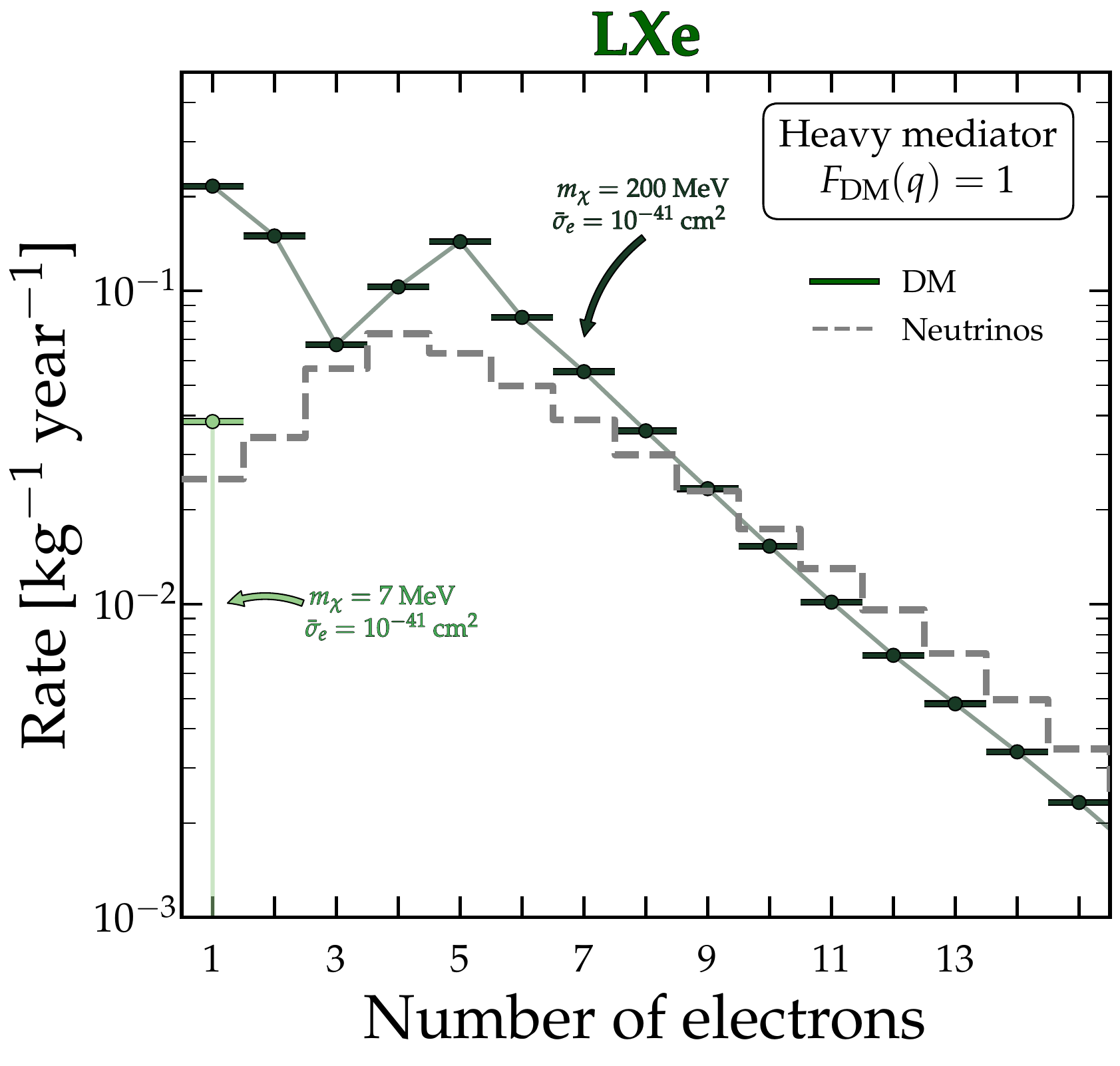}\phantom{0000}
\includegraphics[trim = 0mm 0mm 0mm 0mm, clip, width=0.49\textwidth]{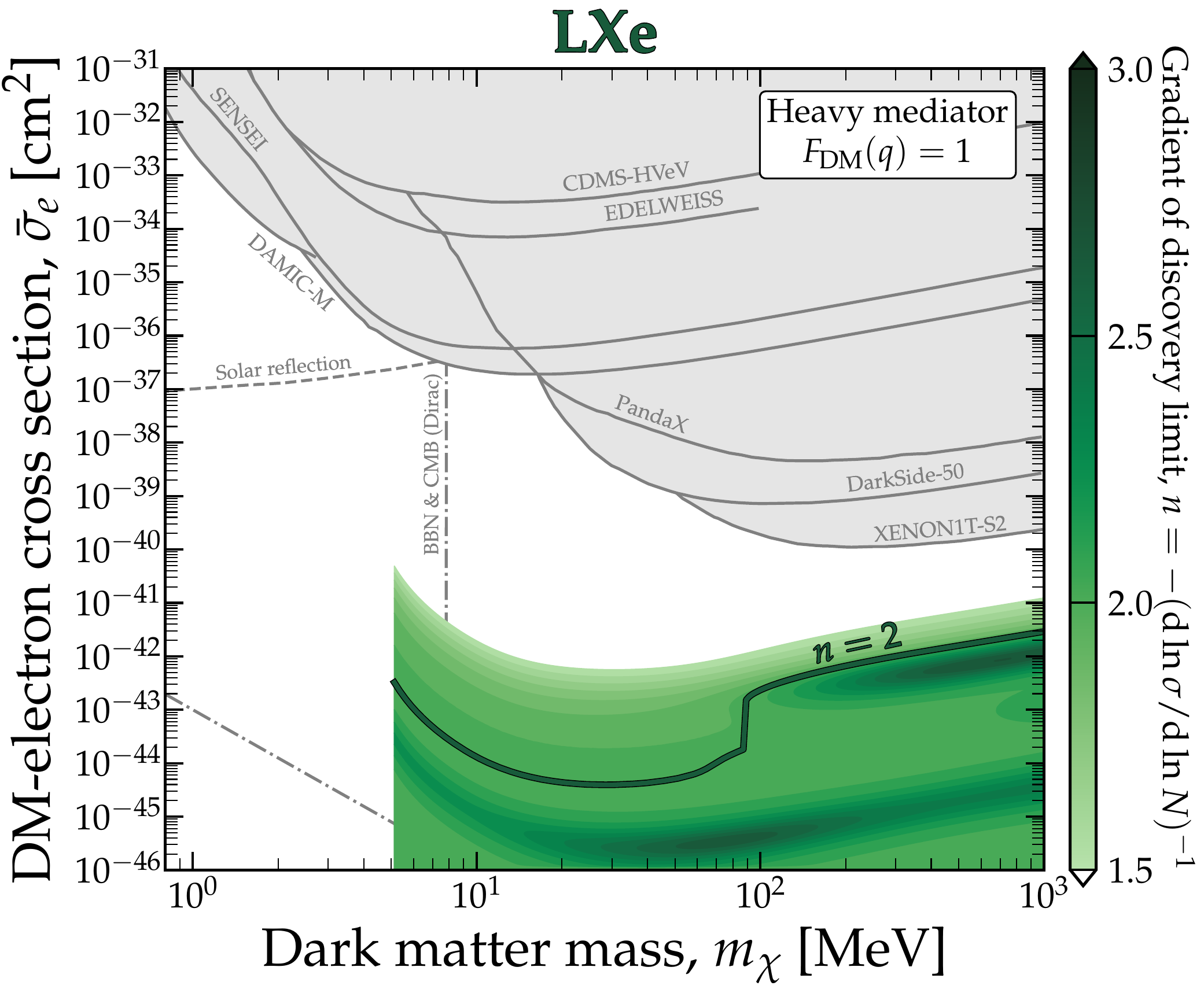}
\caption{{\bf Left panels}: Event rates for DM (solid lines) and neutrino-induced (gray dashed line) electrons for two of the target media studied here---silicon and liquid xenon. In each case, we show two example DM masses assuming a heavy mediator, with cross sections chosen to show cases that have a stronger overlap with the neutrino background. {\bf Right panels}: The corresponding neutrino fogs for the two targets and under the heavy mediator case (light mediator cases are shown in Fig.~\ref{fig:nufogs_additional_light}). The colorscale shows the value of $n$ calculated from the scaling of the DM discovery as a function of exposure, i.e. $\sigma \propto (MT)^{-1/n}$. We also highlight the $n=2$ contour which we showed in Fig.~\ref{fig:Combined}.} 
\label{fig:nufogs}
\end{center}
\end{figure*}

\begin{figure}
\begin{center}
\includegraphics[trim = 0mm 0mm 0mm 0mm, clip, width=0.47\textwidth]{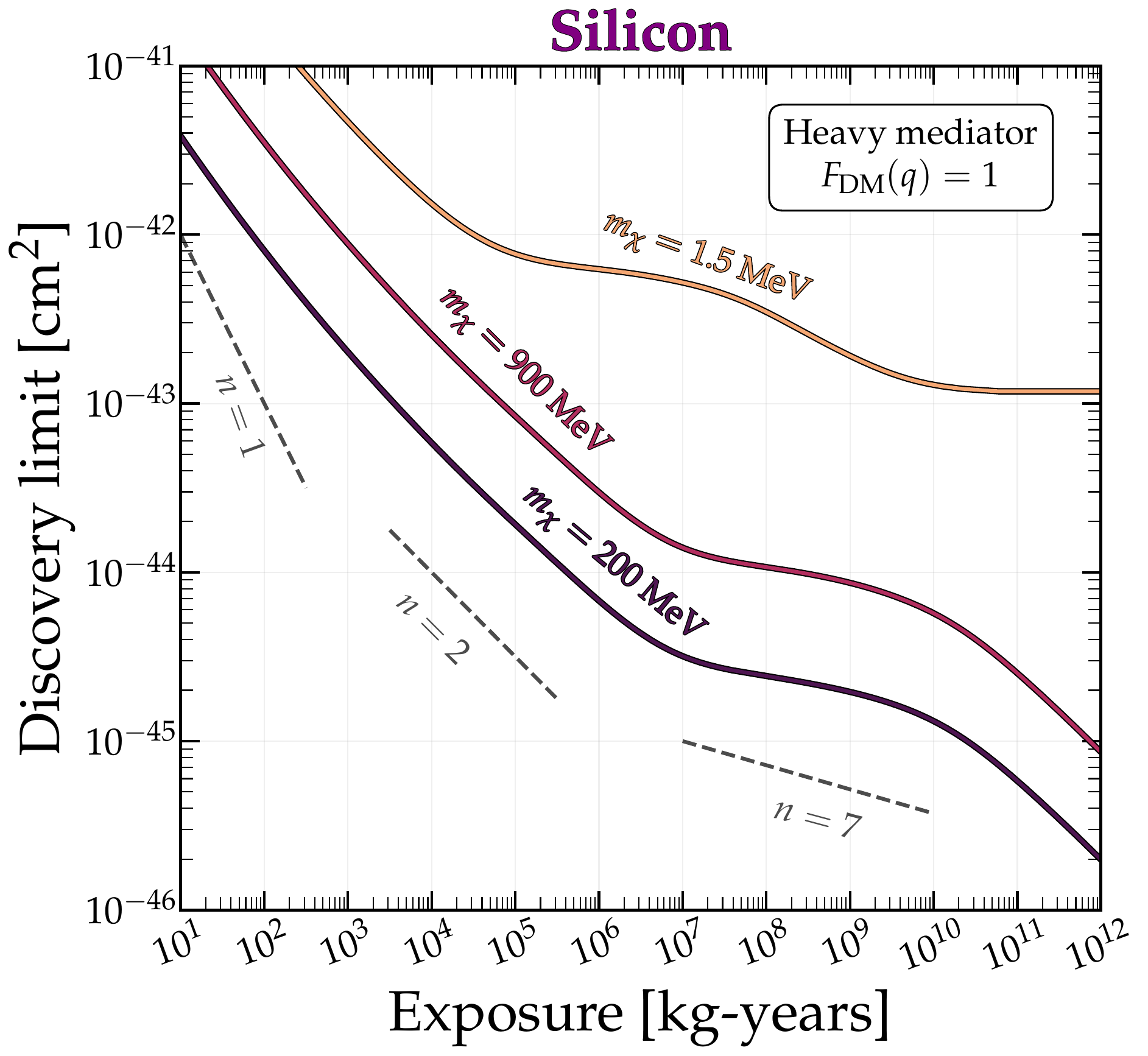}
\caption{Scaling of the DM discovery limit in a silicon target as a function of exposure for three specific DM masses. We see clearly here that there is a hard neutrino \textit{floor} for masses around 1.5 MeV which is a result of the signal being contained in a single electron-number bin. Higher masses do not show this behaviour as expected because the additional spectral information contained over a range of bins allows for greater background discrimination. In those two cases instead, the neutrino fog is less severe, though still present as implied by the values of $n>2$. As a visual aid, we show three straight lines that have gradients of \mbox{$\sigma \propto ({\rm Exposure})^{-1/n}$ with $n = 1,\,2,\,7$}. We emphasise that this plot is to illustrate the impact of the neutrino background---exposures in excess of the ton-year scale are not currently anticipated for this type of experiment.} 
\label{fig:nufog_1D}
\end{center}
\end{figure} 

The statistical procedure to map the neutrino fog was introduced in Ref.\,\cite{OHare:2021utq}. It is based on the conventional profile likelihood ratio test for the existence of a DM signal (parameterised by a cross-section $\sigma$), against the null hypothesis of events due to the neutrino background only (parameterised by a set of fluxes $\mathbf{\Phi}$). The test is performed assuming a fixed DM mass and is then repeated across the DM mass range of interest. 

The event rates are sorted into bins in terms of number of electrons $n_e$, from which we can calculate the binned likelihood $\mathcal{L}(\sigma,\mathbf{\Phi}|\mathcal{M})$ as the product of the Poisson probabilities for the number of events observed in each bin given an expected number of events. Under the alternative hypothesis, $\mathcal{M}$, the expected number of events is the sum of the neutrino-induced events and the DM ones, whereas under the null hypothesis $\mathcal{M}_{\sigma=0}$ the DM-cross section is set to zero and hence we count only neutrino events. Importantly for the test, the null hypothesis is a special case of the alternative hypothesis. We also multiply the likelihood by a set of Gaussian distributions, one for each neutrino source in $\mathbf{\Phi}$, which we treat as nuisance parameters with central values and uncertainties from our neutrino flux model~\cite{OHare:2022jnx}. 

These likelihoods are maximised to construct the profile-likelihood-ratio test statistic,
\begin{align}
q_0 &= 
\begin{cases}
-2\ln{\left[\frac{\mathcal{L}(0,\hat{\hat{\mathbf{\Phi}}}|\mathcal{M}_{\sigma=0})}{\mathcal{L}(\hat{\sigma},\hat{\mathbf{\Phi}}|\mathcal{M})}\right]} & \hat{\sigma} > 0\\
0 & \hat{\sigma} \leq 0 \, .
\end{cases}
\label{eq:TS}
\end{align}
The likelihood $\mathcal{L}$ is maximised at $\hat{\hat{\mathbf{\Phi}}}$ when $\sigma=0$ and at $(\hat{\sigma},\hat{\mathbf{\Phi}})$ when $\sigma$ is allowed to vary. We evaluate $q_0$ using the Asimov dataset approximation, where the observed number of events in each bin is unphysically set to be exactly equal to the expected number. The Asimov approximation is accurate enough in the high-statistics regime while remaining computationally inexpensive. Since the two models in Eq.\,\eqref{eq:TS} differ by only one parameter $\sigma$, in the asymptotic limit, $q_0$ will follow a $\frac{1}{2}\chi^2 + \frac{1}{2} \delta(0)$ distribution when $\mathcal{M}_{\sigma=0}$ is true \cite{Chernoff:1954eli,Cowan:2010js}. This means the significance of the DM signal over the background is simply $\sqrt{q_0}$. When the alternative hypothesis is true, on the other hand, $q_0$ follows a non-central $\chi^2$ distribution (which can also be evaluated analytically, see Eq.~11 of Ref.~\cite{Tang:2023xub}).

In this work, we construct the neutrino fog from median $3\sigma$ discovery limits by finding the smallest DM cross-section for which $q_0>9$ in 50\% of hypothetical experiments---the use of the Asimov dataset ensures that the latter of those two conditions is met. These discovery limits will decrease as the exposure increases, however the \textit{rate} at which they decrease depends on how badly the neutrino background overlaps with the signal for that particular DM mass. This can be seen in Fig.~\ref{fig:nufog_1D} where we show the decrease in the discovery limits for three example DM masses in a silicon target experiment. The fact that these lines do not decrease as expected for Poissonian statistics $\sigma \propto 1/\sqrt{\rm Exposure}$----and instead have temporary plateaus---this is the detrimental effect of the neutrino fog. 

So to quantify the ``opacity'' of neutrino fog, we can just consider the gradient of these lines. As in Ref.~\cite{OHare:2021utq} we calculate this as,
\begin{align}
  n &= - \left(\dv{\ln\sigma}{\ln MT}\right) \, ,
\end{align}
where we express the exposure as the detector mass multiplied by the running time, $MT$. This is, in turn, proportional to the observed number of background events $N$, which we can use interchangeably in this discussion. 

The opacity parameter is defined such that $n=2$ indicates normal Poissonian background subtraction, and any value $n > 2$ indicates that signal discovery power is being suppressed by a background. There are four distinct regimes in the value of $n$ as we increase the exposure from a very small value to a very large one. When the exposure is too small to detect the neutrino background ($N\ll1$), the discovery limit scales as $\sigma \propto 1/N$, i.e.~$n=1$. As the exposure increases however, and background events are detected, the scaling transitions to the expected Poisson regime $\sigma \propto \sqrt{1/N}$, or $n=2$. The next regime occurs when the DM signal lies underneath the neutrino background fluctuations. Here, the discovery limit scales as $\sigma \propto \sqrt{(1+N\delta\mathbf{\Phi}^2)/N}$, or $n>2$ \cite{Billard:2013qya}. Finally, when the number of observed events is so large that the experiment can effectively measure its own background, the Poissonian regime returns. However, we emerge into this final regime only if there is some information present in the observable at hand (in our case, the recoil spectrum), that distinguishes the signal from the background. If the signal and background lead to \textit{identical} observables, then there will be a hard neutrino floor: $n\to \infty$.

To visualise the neutrino fog, we can simply colour-code the DM parameter space by the value of $n$. This leads to the plots in the right-hand panels of Fig.\,\ref{fig:nufogs}. We present the silicon and LXe neutrino fogs as representatives of solid and liquid detectors. The neutrino fogs for the light mediator case, as well as for germanium and LAr, are presented in Figs.\,\ref{fig:nufogs_additional_heavy} and \ref{fig:nufogs_additional_light}. The darker-coloured regions represent areas where $n > 2$, meaning the DM and neutrino event rates have greater spectral degeneracy. In most cases, there are slight differences between the DM and neutrino signals, which make it possible to distinguish the two in the high-statistics regime. Therefore, we find that once $N$ has grown large enough, the scaling returns back to $n=2$ for very low cross sections.

On the other hand, for very low DM masses in the semiconductor and LAr cases, both the DM signal and the low-energy component of the neutrino background ($pep$) are contained only in a single bin in $n_e$. Because of this, we have a hard neutrino \textit{floor}. The reason behind this is related to what we stated above: when the signal and a background component are contained in a single bin, there is no information available to the experiment that could ever distinguish them, even in the limit of a huge exposure. This occurs, for example, around 1.5\,\MeV in the silicon case. Similar features are absent for xenon and argon at higher masses, as none of the solar neutrino fluxes generates \textit{only} single-electron events under our adopted ionisation model. 

Following Ref.~\cite{OHare:2021utq}, we define the cross section at which the gradient departs from $n=2$ to $n>2$ as the neutrino fog ``edge''. The two baseline neutrino fog edges for silicon and xenon are shown in Fig.~\ref{fig:Combined}. We define these merely for convenience and future reference, so that the region of parameter space that is impacted by the neutrino background can be visualised in a more simple way. 

Going back to the first figure, Fig.~\ref{fig:Combined}, we compared our neutrino fog edges with some example DM models and production scenarios: an elastically decoupling relic (ELDER)\,\cite{Kuflik:2015isi, Kuflik:2017iqs}, scalar freeze out\,\cite{Essig:2022dfa} and the DM freeze-in mechanism~\cite{Essig:2012yx, Essig:2022dfa}. These are motivated models that are targeted by the community for future experiments, and interestingly, the former two are partially obscured by the neutrino fog. Fortunately, the freeze-in scenario appears to be safe and testable well before the neutrino background becomes problematic.

\section{Discussion and conclusions}\label{sec:conc}
With the continued interest in sub-\GeV DM detection, it is important to understand the limits that the neutrino background places on our ability to reach smaller cross-sections and lighter masses, and how this might impact the future of the field. In this paper, we have attempted to showcase these limits visually by mapping the so-called neutrino fog.

Oscura is the name of the planned next-generation silicon-based skipper-CCD detector. A solar neutrino signal is expected to be detected by an experiment at this scale, but the projected sensitivity limits do not quite reach the edge of the neutrino fog as we have derived it here. Oscura should reach $\bar{\sigma}_e = 10^{-43}$ cm$^2$ for masses $m_\chi \sim 10$ \MeV, assuming a 30 kg-year exposure\,\cite{Oscura:2022vmi}. Therefore, its sensitivity will barely reach the edge of the fog, but the sharp increase in opacity below the floor renders greater exposures futile anyway---DM models with masses around 1 MeV and cross sections below $10^{-42}$~cm$^2$ are essentially undiscoverable in any type of detector proposed to date.

While neutrinos seem to be far away from threatening solid-state experiments, for liquid argon (shown in the appendix\,\ref{app:other_fog}) and, to some extent, liquid xenon, the onset of the fog emerges at much higher cross sections and so could potentially impact long-term projects like XLZD. The neutrino fog will also impact our ability to probe the theoretically-motivated regions of the sub-\GeV parameter space. If DM interacts through a heavy mediator, the window of theoretically motivated DM models is obscured by the fog above $m_\chi \sim 100$ \MeV for all targets we study here. In the case of a light mediator, the prime theory target of freeze-in DM is not obscured by the neutrino fog, and so the neutrino background will remain largely irrelevant for its discovery or exclusion.

Before finishing, it is interesting to ponder ways in which the neutrino fog may be circumvented if the community decides it is desirable to push the search for DM-electron interactions down to cross sections as low as the ones we have studied here. Going \textit{through} the neutrino fog essentially requires that we provide experiments with additional information to distinguish the DM signal from its background. A few ideas have been presented in the past, including annual modulation~\cite{Davis:2014ama}, directional detection~\cite{O'Hare:2015mda, Grothaus:2014hja, Mayet:2016zxu, OHare:2017rag, Franarin:2016ppr, OHare:2020lva, Vahsen:2020pzb, Vahsen:2021gnb, OHare:2022jnx}, as well as improving our prior knowledge of neutrino flux normalisations~\cite{OHare:2020lva, OHare:2021utq, Ruppin:2014bra}. 

Unfortunately, the annual modulation of the solar neutrino and DM fluxes are at the percent level and so are too small to provide a practical means to push through the neutrino fog. Directionality also seems additionally challenging for DM-electron scattering, though some ideas exist in the literature, for example detectors composed of graphene layers~\cite{Hochberg:2016ntt, Wang:2015kya, Kim:2020bwm, Catena:2023awl, Catena:2023qkj, Das:2023cbv} or detectors using materials that exhibit anisotropic responses to recoil events~\cite{Blanco:2019lrf, Boyd:2022tcn}. The simplest approach, then, seems to be to just wait for the improvements in our knowledge of neutrino fluxes to be provided by the upcoming generation of large neutrino observatories.

Interestingly, in the case of the neutrino fog we have studied here, another possibility for pushing through it arises. Throughout our study, we have assumed that the primary source of neutrino events is actually the misidentified electron recoils arising from \cevns, and not from direct neutrino-electron interactions, which lead to a comparatively small number of events in the energy range of interest. This means that if experiments can discriminate nuclear and electron recoils, the neutrino fog could be alleviated further. Unfortunately, this is not currently possible at such low energies, where the events are already hard enough to detect in the first place. Some detectors, like dual-phase liquid-noble TPCs, do have methods to distinguish electron recoils from nuclear recoils, but to achieve sensitivity at such low energies, they currently must sacrifice their primary scintillation signal, which is what provides the information required to discriminate electrons from nuclei. Nonetheless, if a detection scheme were devised in which recoiling nuclei and electrons are distinguishable even down to the level of single-to-few primary ionised electrons, then such an experiment could be foreseen to have a much less severe neutrino fog.

\noindent \textbf{\textit{Acknowledgements}}.---This work was supported by The University of Sydney and the ARC Centre of Excellence for Dark Matter Particle Physics. TNM would like to acknowledge the financial support from the Department of Atomic Energy, Government of India, for the Regional Centre for Accelerator-based Particle Physics (RECAPP), Harish-Chandra Research Institute. CAJO is supported by the ARC under grant number DE220100225. We thank B.~M.~Roberts, Tanner Trickle, and Shawn Westerdale for discussions.

\appendix

\section{Additional neutrino fog visualisations}
\label{app:other_fog}
\begin{figure*}[htb]
\begin{center}
\includegraphics[trim = 0mm 0mm 0mm 0mm, clip, width=0.45\textwidth]{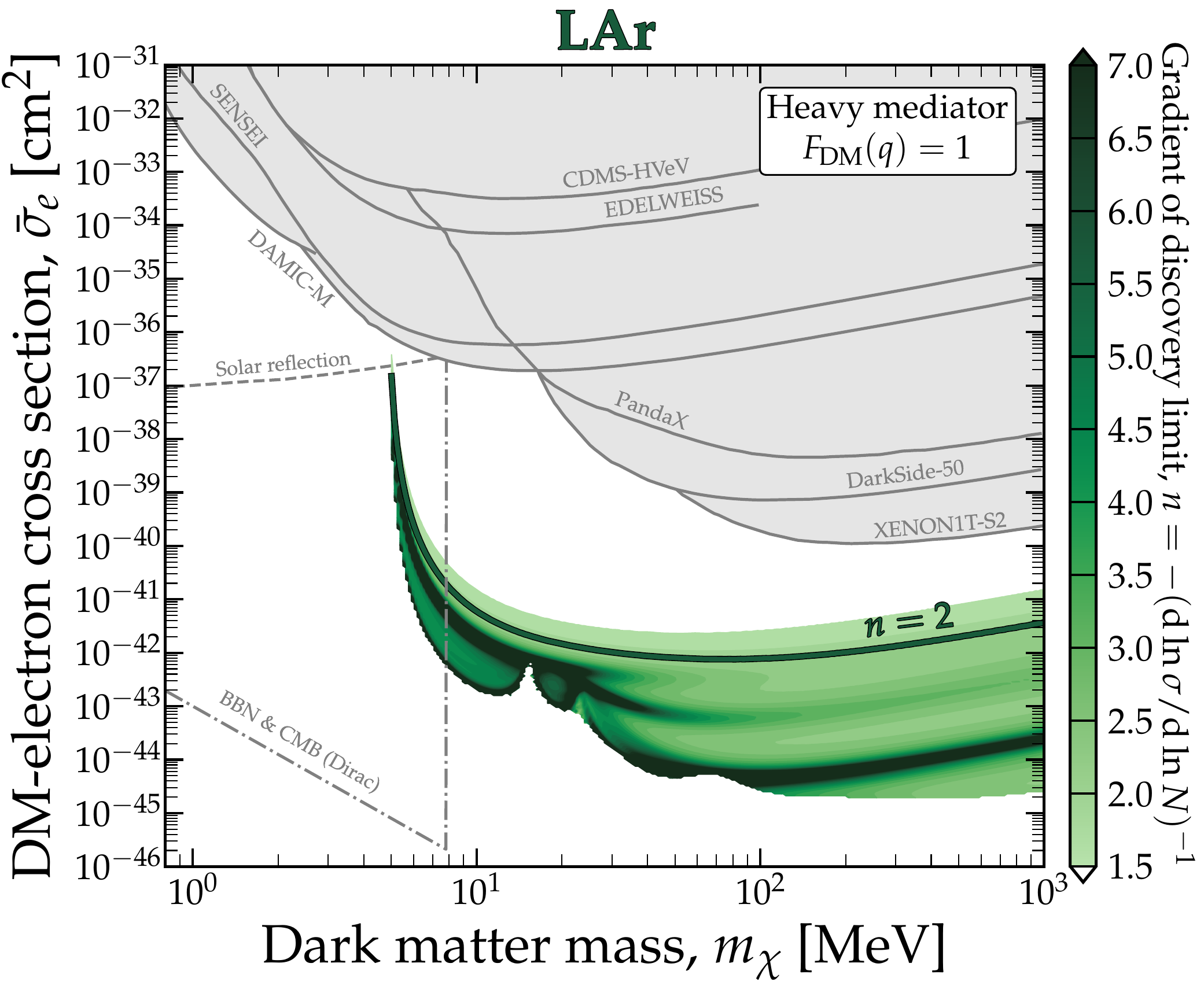}\hspace{3em}
\includegraphics[trim = 0mm 0mm 0mm 0mm, clip, width=0.45\textwidth]{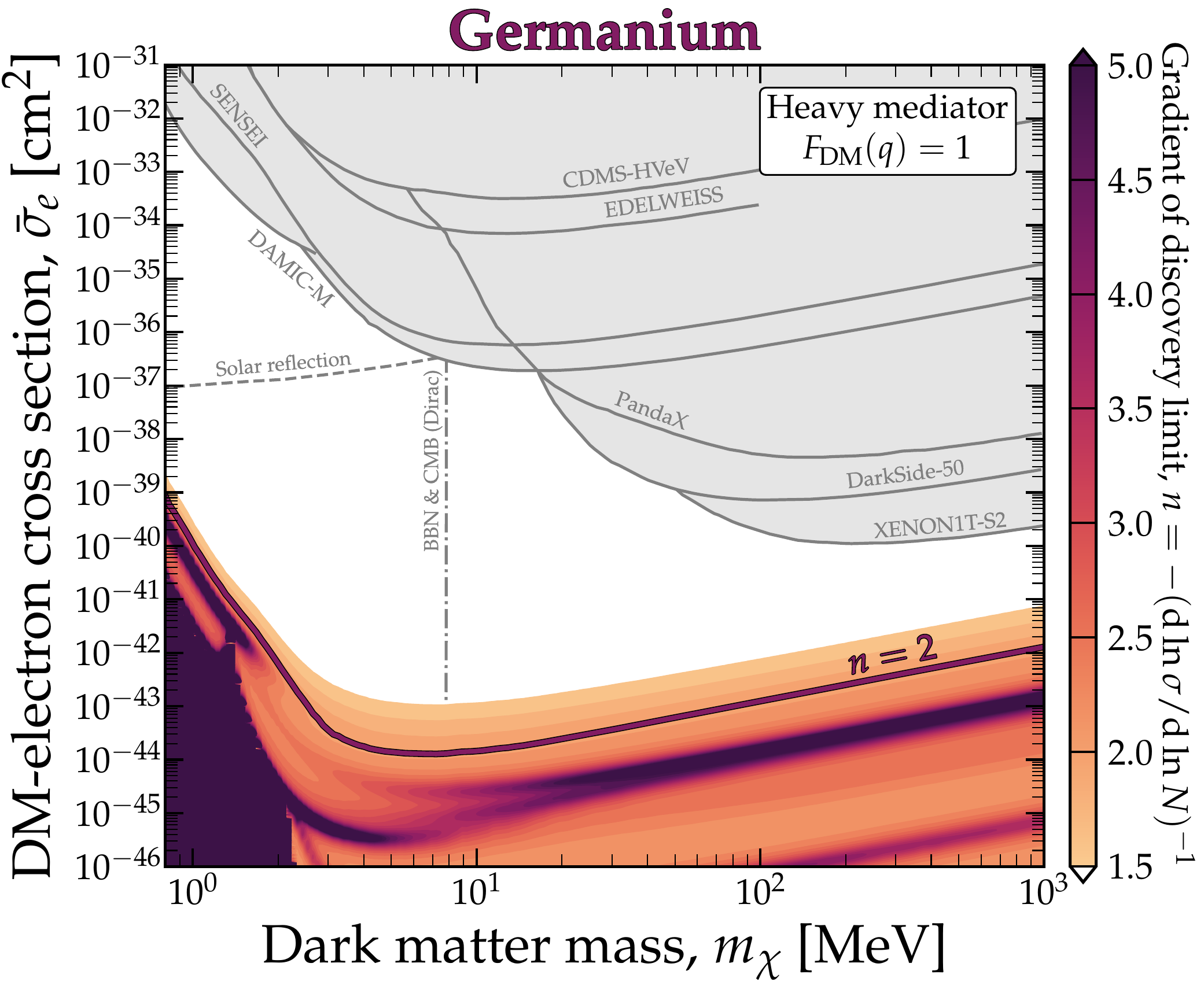}
\caption{Additional neutrino fog visualisations for argon and germanium targets assuming a heavy mediator. For Argon, the required exposures to map the lower extent of the plot are in excess of $10^{15}$ kg-years, leading to numerical instability in deriving an accurate value of $n$. In this case, we simply leave this region unfilled.} 
\label{fig:nufogs_additional_heavy}
\end{center}
\end{figure*}

\begin{figure*}[htb]
\begin{center}
\includegraphics[trim = 0mm 0mm 0mm 0mm, clip, width=0.45\textwidth]{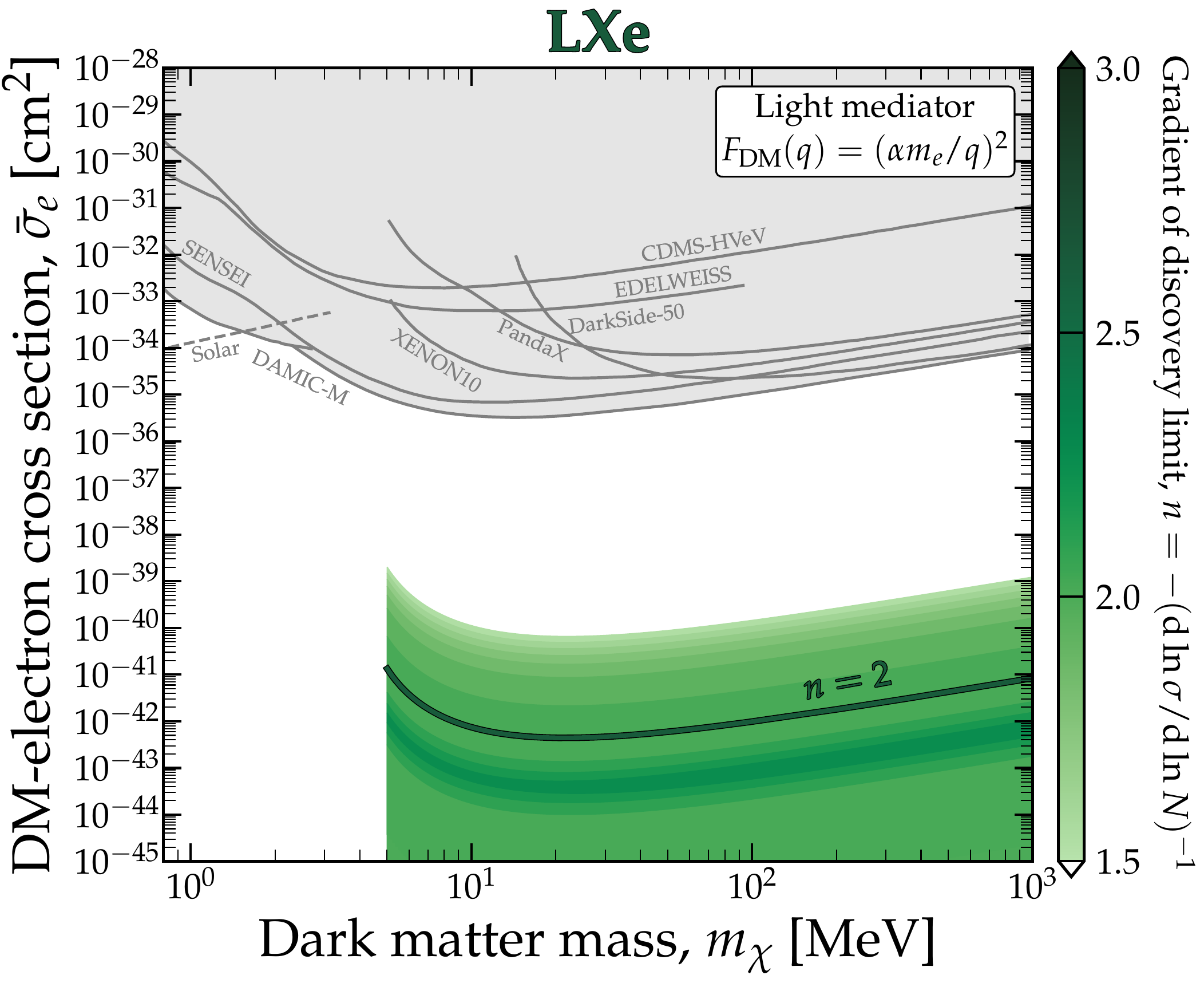}\hspace{3em}
\includegraphics[trim = 0mm 0mm 0mm 0mm, clip, width=0.45\textwidth]{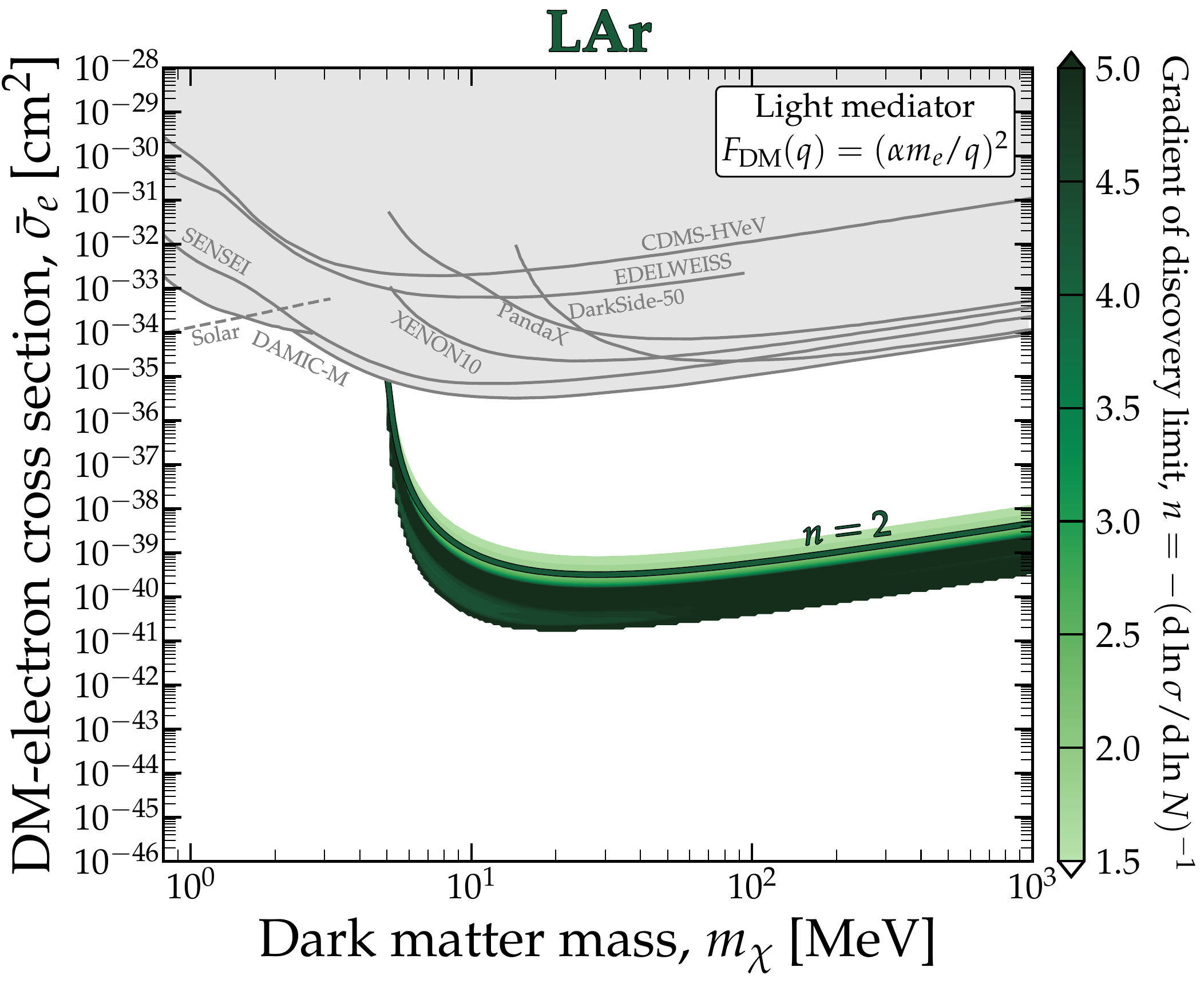}
\includegraphics[trim = 0mm 0mm 0mm 0mm, clip, width=0.45\textwidth]{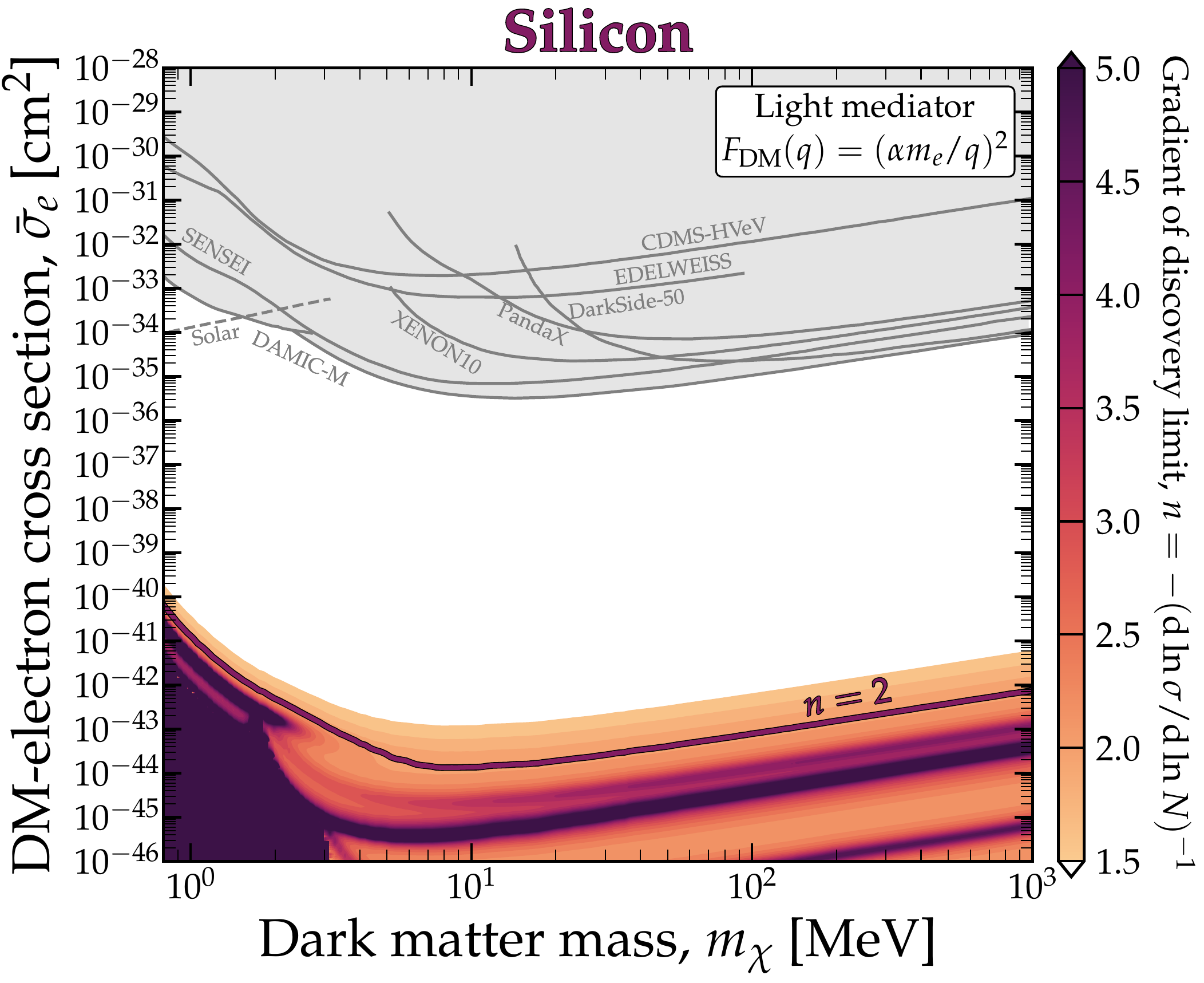}\hspace{3em}
\includegraphics[trim = 0mm 0mm 0mm 0mm, clip, width=0.45\textwidth]{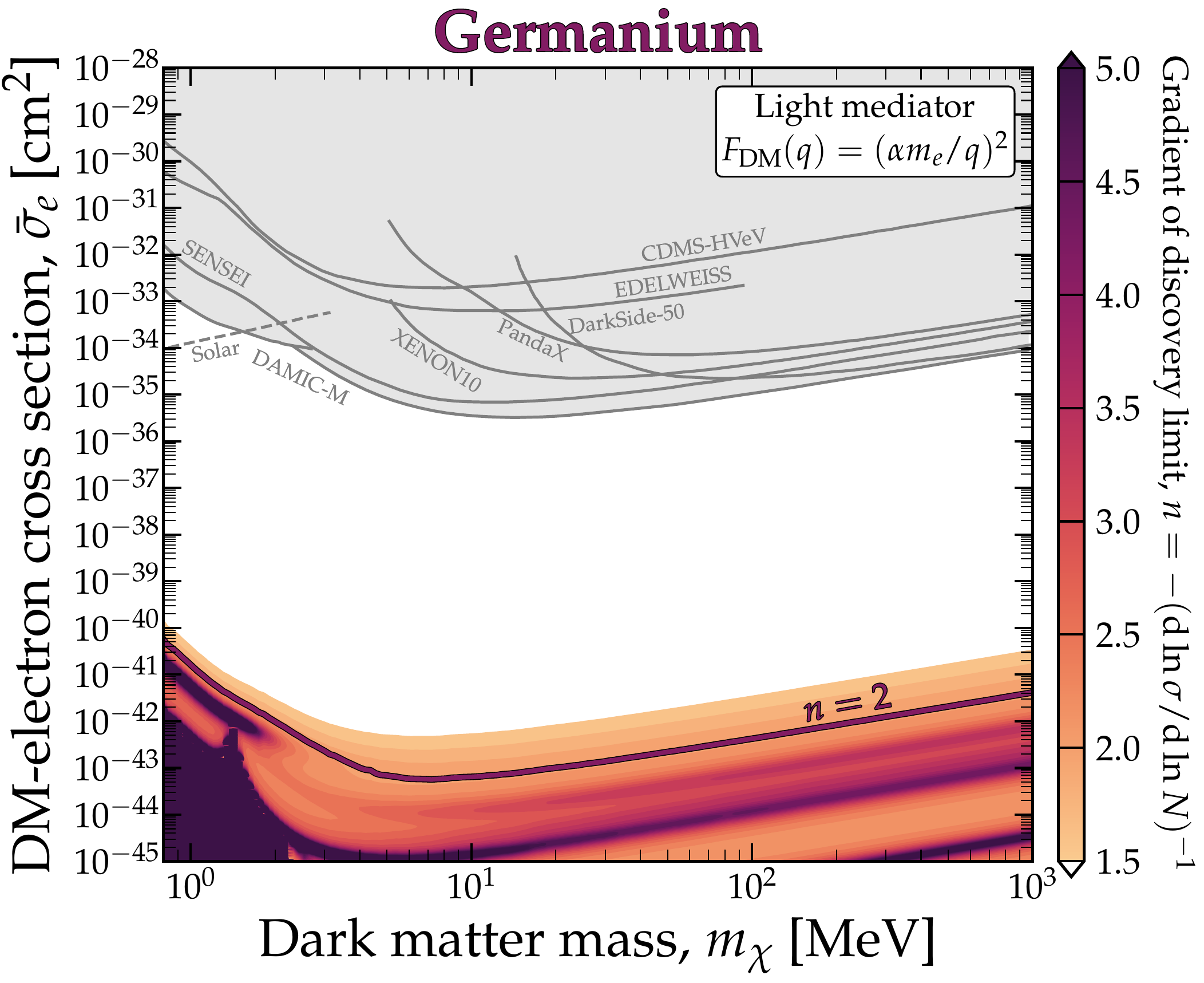}
\caption{Additional neutrino fog visualisations for the light mediator case. For Argon, the required exposures to map the lower extent of the plot are in excess of $10^{15}$ kg-years, leading to numerical instability in deriving an accurate value of $n$. In this case, we simply leave this region unfilled.} 
\label{fig:nufogs_additional_light}
\end{center}
\end{figure*}

In this appendix, we include several further visualisations of the neutrino fog that were not included in the main text. These include the argon and germanium neutrino fogs for the heavy-mediator case (Fig.~\ref{fig:nufogs_additional_heavy}), and all four targets assuming a light mediator (Fig.~\ref{fig:nufogs_additional_light}).

\section{The edge of the neutrino fog using other codes}
\label{app:other_codes}
\begin{figure*}[htb]
\begin{center}
\includegraphics[trim = 0mm 0mm 0mm 0mm, clip, width=0.45\textwidth]{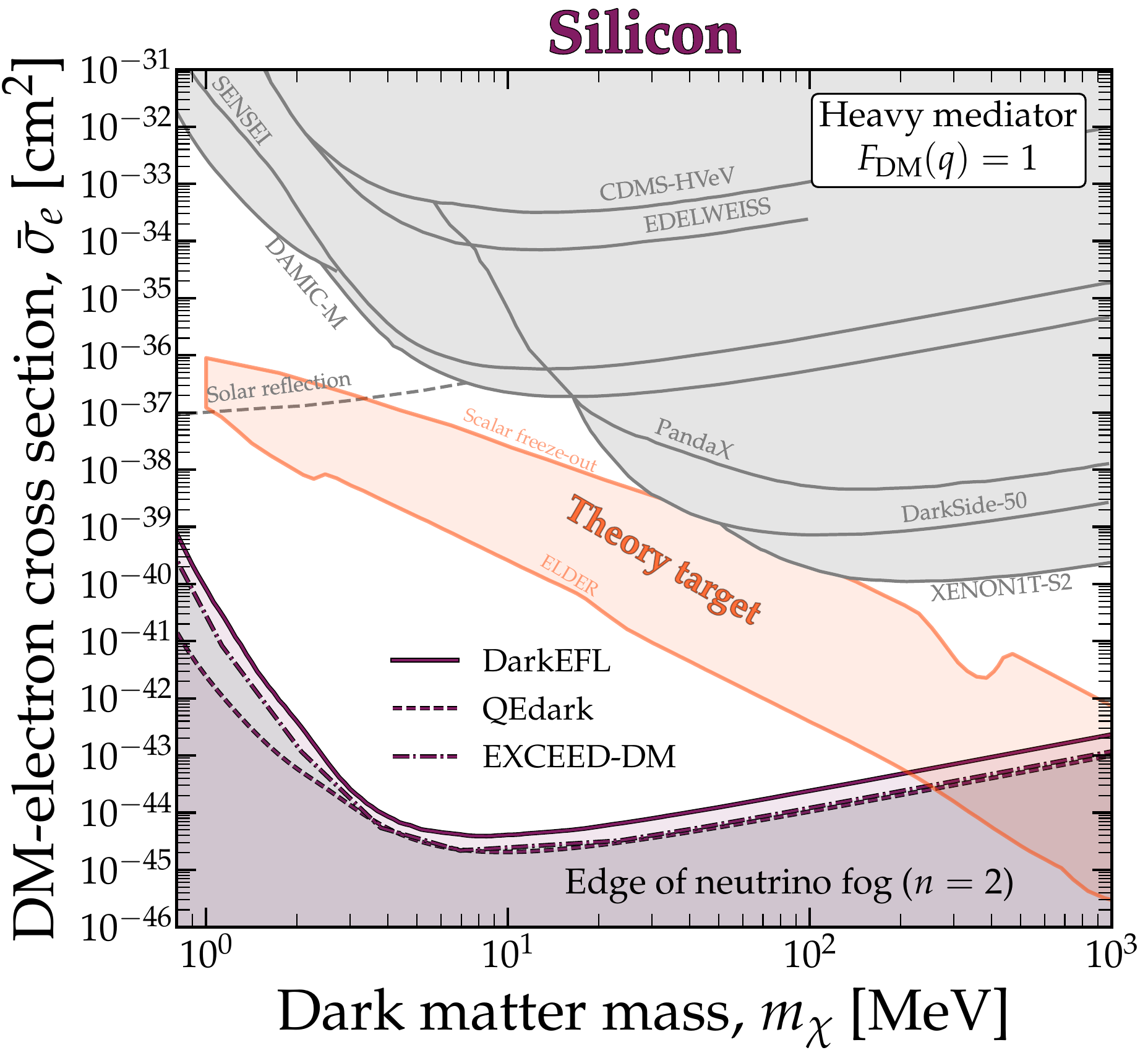}\hspace{3em}
\includegraphics[trim = 0mm 0mm 0mm 0mm, clip, width=0.45\textwidth]{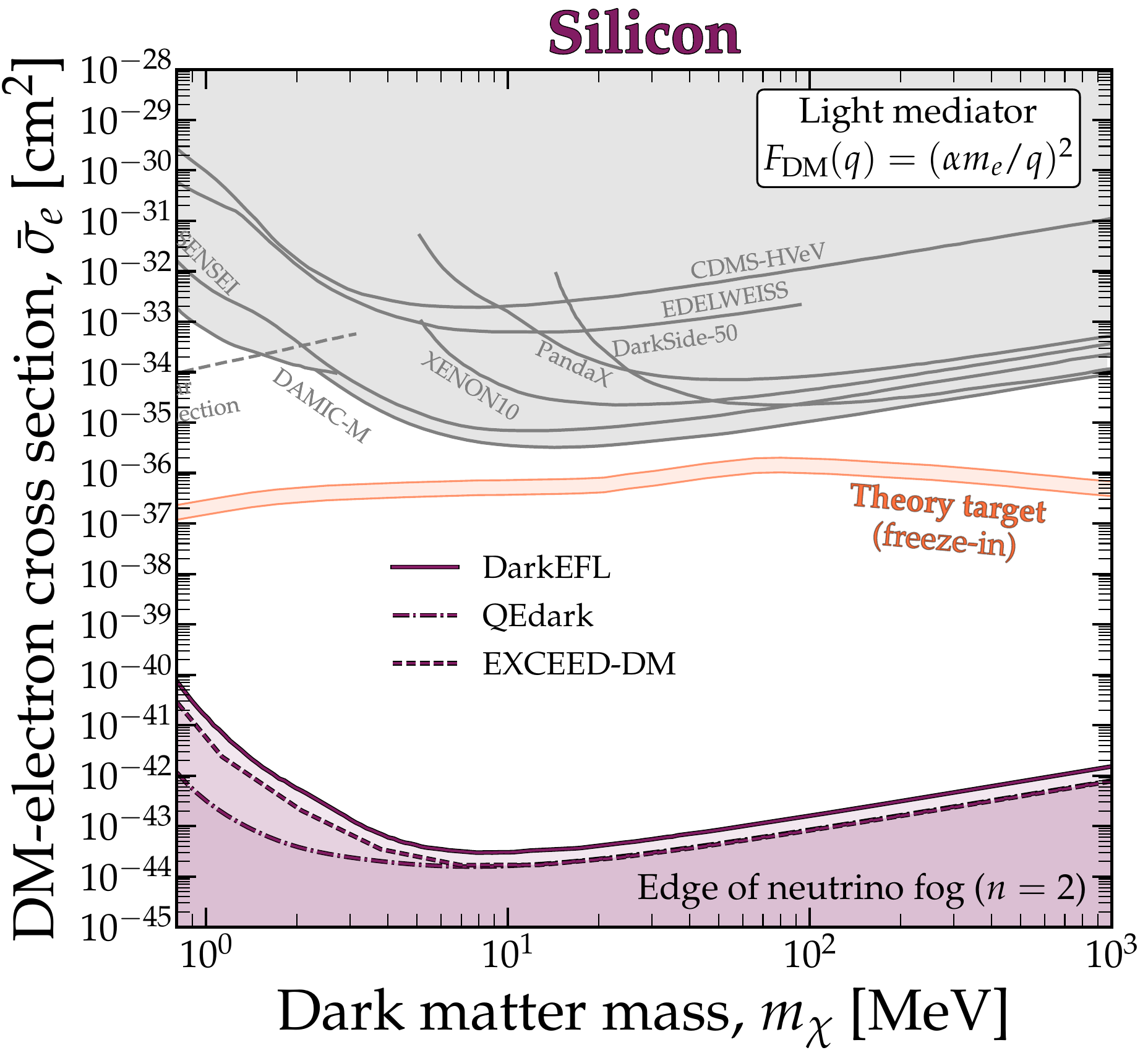}
\caption{The $n=2$ boundaries of the neutrino fog when DM-electron scattering are calculated using DarkEFL (as utilized throughout the paper)\,\cite{Knapen:2021bwg}, QEdark\,\cite{Essig:2015cda}, and EXCEED-DM\,\cite{Trickle:2022fwt} for Si target.} 
\label{fig:nufogs_other}
\end{center}
\end{figure*}

In Fig.\,\ref{fig:nufogs_other}, we compare the $n=2$ edge of the neutrino fog for the cases where DM-electron ionization probabilities are calculated using other publicly available codes QEdark\,\cite{Essig:2015cda}, and EXCEED-DM\,\cite{Trickle:2022fwt}. The result of QCdark\,\cite{Dreyer:2023ovn} agrees well with EXCEED-DM; thus, the corresponding neutrino fog is expected to reside close to the same of the latter. While we have presented our results for Si in Fig.\,\ref{fig:nufogs_other}, we observe similar deviations for the Ge target as well.  

\bibliography{neutrinos.bib}
\bibliographystyle{bibi}

\end{document}